\documentclass[fleqn,usenatbib]{mnras}

\usepackage{newtxtext,newtxmath}

\usepackage[T1]{fontenc}
\usepackage{ae,aecompl}

\usepackage{graphicx}	
\usepackage{amsmath}	

\newcommand{\rc}{\color{black}}

\title[Lyman-$\alpha$ transits]{The fundamentals of Lyman-$\alpha$ exoplanet transits}

\author[Owen, J. E. et al.]{James E. Owen$^{1}$\thanks{E-mail: james.owen@imperial.ac.uk}, Ruth A. Murray-Clay$^{2}$, Ethan Schreyer$^{1}$, Hilke E. Schlichting$^{3,4,5}$, \newauthor David Ardila$^{6}$, Akash Gupta$^{3}$, R. O. Parke Loyd$^{7,8}$, Evgenya L. Shkolnik$^{7}$, \newauthor David K. Sing$^{9,10}$ and Mark R. Swain$^{6}$ \\
$^{1}$Astrophysics Group, Imperial College London, Blackett Laboratory, Prince Consort Road, London SW7 2AZ, UK\\
$^{2}$ Department of Astronomy and Astrophysics, University of California, Santa Cruz, CA 95064, USA\\
$^{3}$Department of Earth, Planetary, and Space Sciences, University of California, Los Angeles, CA 90095, USA\\
$^{4}$Department of Physics and Astronomy, University of California, Los Angeles, CA 90095, USA\\
$^{5}$Department of Earth, Atmospheric and Planetary Sciences, Massachusetts Institute of Technology, MA 02139, USA\\
$^{6}$Jet Propulsion Laboratory, California Institute of Technology, 4800 Oak Grove Drive, Pasadena, California 91109, USA\\
$^{7}$School of Earth and Space Exploration, Arizona State University, Tempe, AZ 85287, USA\\
$^{8}$Eureka Scientific, 2452 Delmer Street Suite 100, Oakland, CA, 94602-3017, USA\\
$^{9}$Department of Earth \& Planetary Sciences, Johns Hopkins University, Baltimore, MD, USA\\
$^{10}$Department of Physics \& Astronomy, Johns Hopkins University, Baltimore, MD, USA}

\date{Accepted XXX. Received YYY; in original form ZZZ}

\pubyear{2015}

\begin{document}
\label{firstpage}
\pagerange{\pageref{firstpage}--\pageref{lastpage}}
\maketitle

\begin{abstract}
Lyman-$\alpha$ transits have been detected from several nearby exoplanets and are one of our best insights into the atmospheric escape process. However, due to ISM absorption, we typically only observe the transit signature in the blue-wing, making them challenging to interpret. This challenge has been recently highlighted by non-detections from planets thought to be undergoing vigorous escape. Pioneering 3D simulations have shown that escaping hydrogen is shaped into a cometary tail receding from the planet. Motivated by this work, we develop a simple model to interpret Lyman-$\alpha$ transits. Using this framework, we show that the Lyman-$\alpha$ transit depth is primarily controlled by the properties of the stellar tidal field rather than details of the escape process. Instead, the transit duration provides a direct measurement of the velocity of the planetary outflow. This result arises because the underlying physics is the distance a neutral hydrogen atom can travel before it is photoionized in the outflow. Thus, higher irradiation levels, expected to drive more powerful outflows, produce weaker, shorter Lyman-$\alpha$ transits because the outflowing gas is ionized more quickly. Our framework suggests that the generation of energetic neutral atoms may dominate the transit signature early, but the acceleration of planetary material produces long tails. Thus, Lyman-$\alpha$ transits do not primarily probe the mass-loss rates. Instead, they inform us about the velocity at which the escape mechanism is ejecting material from the planet, providing a clean test of predictions from atmospheric escape models. 
\end{abstract}




\section{Introduction}
Many exoplanets reside in close proximity to their host stars, and it is now well established that the majority of low-mass stars host at least one, and in many cases several planets with orbital periods $\lesssim 100~$days \citep[e.g.][]{Fressin2013,Dressing2015,Silburt2015,Mulders2018,Zink2019,Zhu2021}. Since these planets reside so close to their parent stars they experience extreme levels of irradiation, which is able to drive vigorous hydrodynamic escape from their atmospheres \citep[e.g.][]{Lammer2003,Yelle2004,GarciaMunoz2007,MurrayClay2009,Owen2012,Ginzburg2018,Kubyshkina2018,Howe2019}.

Over these planet's billion-year lifetimes mass-loss from their primordial hydrogen-dominated atmospheres is one of their dominant evolutionary drivers \citep[e.g.][]{Owen2019,Bean2021}. For example, both the hot-Neptune ``desert'' \citep[e.g.][]{Beauge2013,Mazeh2016,Lundkvist2016} and the radius-valley, where planets with radii of $\sim1.5-2.0$~R$_\oplus$ are less frequent \citep[e.g.][]{Fulton2017,Fulton2018,VanEylen2018} {\rc might be} features carved by atmospheric escape. In-fact both this ``desert'' and radius-valley were  fundamental early predictions of the escape-driven evolutionary model of close-in exoplanets \citep[e.g.][]{Owen2013,Lopez2013,Jin2014,Owen2018,Ginzburg2018}. 

{\rc Comparisons between  exoplanet demographics and  evolutionary models indicate agreement so far at the population level \citep[e.g.][]{Owen2017,Wu2018,Gupta2019,Gupta2020,Rogers2021,Rogers2022}; although evidence of individual systems that don't fully match are starting to emerge \citep[e.g.][]{OCE20,Diamond-Lowe2022}}.  However, this agreement can be achieved using {\rc distinct atmospheric escape models, such as core-powered mass-loss and photoevaporation which predict different mass-loss timescales \citep[e.g.][]{Owen2017,Ginzburg2018}}. Specifically, the ``photoevaporation'' model where mass-loss is driven by heating from stellar extreme ultraviolet (EUV, 100-912~\AA) and X-ray ($< 100$~\AA) photons gives rise to more powerful but shorter-lived mass-loss; while core-powered mass-loss, where escape is driven by the planet's own internal luminosity and stellar bolometric luminosity results in weaker but longer-lived mass-loss \citep[see][for a recent review]{Bean2021}. While different atmospheric escape models do impact the exoplanet population in subtly different ways \citep[e.g.][]{Rogers2021b}, such demographic tests require a much larger number of observed exoplanets than are known today. This agreement between models and observations excludes the possibility that atmospheric escape is unimportant, and the exoplanet population simply formed in a similar state to that we observe today \citep[e.g.][]{Zeng2019}, or is an imprint of gas accretion \citep[e.g.][]{Lee2021,Lee2022}.  Additionally, theoretical work has explored the possibility that hydrogen from the primordial envelope could be dissolved in the mantle and then subsequently degassed if the primordial envelope is lost \citep{Chachan2018}, although there is debate about how efficient this process is \citep{Schlichting2021}. If a significant amount of hydrogen ($\gtrsim1\%$ by mass) is degassed from the mantle and subsequently lost, it constitutes an additional component to be considered when modelling how atmospheric loss may sculpt the exoplanet occurrence rate as a function of radius \citep{Kite2019}.  Though different atmospheric escape models can be used to reproduce observed planetary demographics, these models have different implications for the formation history, chemical evolution, and current chemistry of observed planets, so finding alternative ways to distinguish between them is important.

In-order to quantitatively test our atmospheric escape and planetary evolution models we require direct observations of ongoing mass-loss that can be tested against various predictions. Currently, ongoing atmospheric escape has been observed using transmission spectroscopy in the HeI 10830~\AA~ line \citep[e.g.][]{Spake2018,Allart2018}, H$\alpha$ \citep[e.g.][]{Jensen2012,Cauley2017} and UV metal lines \citep[e.g.][]{Vidal-Madhar2004,Loyd2014,Loyd2017,dosSantos2019,Sing2019,BenJaffel2022}. Yet the most successful observational probe of escape is through Lyman-$\alpha$ transmission spectroscopy with detections from hot-Jupiters HD 209458 b \citep{Vidal-Madjar2003} and HD 189733 b \citep{Lecavelier2010,Lecavelier2012}, as well as warm Neptunes GJ 436 b \citep{Kulow2014,Ehrenreich2015,Lavie2017} and GJ 3470 b \citep{Bourrier2018}.  Lyman-$\alpha$ transits are not a perfect probe, there are several curious non-detections, including from the highly irradiated sub-Neptune planets HD 97658 b \citep{HD97658Ly1}, $\pi$ Men c \citep{GarciaMunoz2020} and the young planet K2-25 b \citep[e.g.][]{Rockcliffe2021}, yet large tentative transits from the more weakly irradiated sub-Neptunes K2-18b \citep{dosSantos2020}. Even within the same system, the young star HD 63433, planet b shows no evidence for a Lyman-$\alpha$ transit, while planet c shows a significant transit \citep{Zhang2021}. Additionally, interstellar absorption and geocoronal emission obscure the line-core in the velocity range of approximately $-50$ to $50$~km~s$^{-1}$ \citep[e.g.][]{Landsman1993,Wood2005}. As the typical velocity for gas undergoing atmospheric escape is $\lesssim 20$~km~s$^{-1}$ when it leaves the planet's Hill sphere, the lack of accessibility of the line-core means Lyman-$\alpha$ transits cannot be used to probe the launching region of the wind (typically withing several planetary radii), unless the host star has a high radial velocity relative to the ISM. In fact the Lyman-$\alpha$ absorption we observe is typically {\rc Doppler-shifted to velocities higher than the typical thermal velocities of planetary outflows ($\lesssim 10$~km~s$^{-1}$), including absorption at velocities $\sim -50$ to $-100$~km~s$^{-1}$ indicating the outflow is interacting with the circumstellar environment}.  However, why we observe Lyman-$\alpha$ transits around some planets and not others, and the mechanism that gives rise to large blue-shifts is poorly understood. The non-detections around young planets are particularly important to understand because they are large enough that they must host hydrogen dominated atmospheres \citep{Owen2020}, and are thought to be experiencing the majority of their mass-loss at these ages. {\rc It is worth noting that the non-detections from HST are of variable constraining power; for example HD 63433b has a 2$\sigma$ upper limit on the transit depth of $\sim 3$\% \citep{Zhang2021}, whereas K2-25 b has a 2$\sigma$ upper-limit of $\sim 30$\% \citep{Rockcliffe2021}. Therefore, some planets may present Lyman-$\alpha$ transit depths of a few-percent, but they remain undetectable due to HST's limited sensitivity.} 

In recent years, there has been significant simulation effort focusing on the Lyman-$\alpha$ transits. \citet{Bourrier2013} extended a 3D particle simulation to model a number of Lyman-$\alpha$ transits \citep[e.g.][]{Bourrier2015,Bourrier2016}. Additionally, there have been a number of 3D hydrodynamic/radiation-hydrodynamic simulations of atmospheric escape \citep[e.g.][]{Bisikalo2013,Matsakos2015,CarrollNellenback2017,Villarreal2018,Esquivel2019,Khodachenko2019,Debrecht2018,McCann2019,Debrecht2020,Harbach2021,Carolan2021,MacLeod2021,Mitani2022} focusing on the interaction between the planetary outflow and the circumstellar environment. These simulations indicate that the gas escaping from the planet is shaped into a cometary tail by the stellar tidal field, ram-pressure from the stellar wind and radiation pressure. Indeed such tail's explain the extended Lyman-$\alpha$ transits of  GJ~436~b, GJ~3470 b. In addition, some works have studied charge-exchange with stellar wind protons yielding fast moving neutral hydrogen (Energetic Neutral Atoms) that provide additional Lyman-$\alpha$ absorption at high-velocities \citep[e.g.][]{Holmstrom2008,Tremblin2013}. Some authors have computed synthetic Lyman-$\alpha$ transits from their simulations, demonstrating qualitative agreement with those that are observed \citep[e.g.][]{Villarreal2021}, indicating the picture of a cometary tail receding from the planet is, at least, qualitatively correct. However, as yet, there is no physical description of Lyman-$\alpha$ transits and how they vary with system parameters such as planet properties, irradiation-levels and mass-loss rates. Further, if the Lyman-$\alpha$ signal primarily arises from Energetic Neutral Atoms we actually only gain insight into the size of the planetary outflow, with the transit signature controlled by the stellar wind properties instead \citep{Holmstrom2008}. All that is currently clear is that ionization by stellar EUV photons, radial acceleration of the escaping planetary gas and charge exchange with the stellar wind all play a role. In particular, there is a limited understanding of how these processes combine such that some planets present large transits (at the $\sim$ 50\% level) while others for seemingly similar planetary types not observed (at the $\lesssim$ 10\% level). 

\section{Overview and Motivation}

There has been significant work developing {\rc analytical and numerical} models of optical photometric exoplanet transits \citep[e.g.][]{Sackett1999,Mandel2002,Seager2003} and the planetary/stellar parameters they encode (such as planet size), but this has not been extended to escaping atmospheres. In this work we develop a {\rc toy} physical picture for a Lyman-$\alpha$ transit to understand its dependence on relevant planetary/stellar parameters and show how this can lead to either large or undetectable transits. In doing so we show that Lyman-$\alpha$ transits can be used to quantitatively probe, and distinguish between mass-loss models.  

The aim of this work is not to replace simulations, quite the contrary, but rather to provide a framework for interpreting both observations and simulations. Multi-dimensional hydrodynamic simulations with radiative transfer (and possibly MHD) are necessary for direct comparison to observations; however, such simulations are computationally challenging and are at the cutting edge of what is possible. This means parameter studies with such simulations are not feasible, and trends are difficult to extract. Currently, simulations do not always provide a good match to the observations. For example, the simulations presented by \citet{Khodachenko2019} are in good agreement with GJ 436 b's transit. {\rc Simulations by \citet{Shaikhislamov2020} can match the non-detection of $\pi$ Men c, but these simulations required either a weaker than estimated stellar wind or higher than estimated ionizing flux. Both these parameters are highly uncertain, and typically estimated via scaling relations (for the stellar wind properties), or extrapolations from other parts of the stellar spectrum (for the ionizing flux).    Simulations presented in \citet{Zhang2021} fail to match the non-detected transits of  HD 63433 b for their estimated ionizing flux and a range of stellar wind properties. Thus, it's unclear if one should interpret these observations as evidence for hydrogen poor sub-Neptunes (as suggested by \citealt{GarciaMunoz2020} and \citealt{Zhang2021}). Alternatively the star could have different wind/ionizing fluxes than used in the modelling of $\pi$ Men c and HD 63433 b or there's missing physics in the simulations. }

However, what these simulations have revealed is that there are several physical processes that play an important role in determining the observational properties of Lyman-$\alpha$ transits. The stellar EUV field not only launches a wind in the case of photoevaporation, but also ionizes neutral hydrogen atoms leaving the planet, rendering them unobservable in Lyman-$\alpha$ \citep{Bourrier2013,Bourrier2016}. Simulations by \citet{Holmstrom2008,Tremblin2013,Lavie2017,Kislyakova2019,Khodachenko2019} have noted the importance of charge-exchange in producing energetic neutral atoms (ENAs) that could be observed through their transit signature in the blue-wing of Lyman-$\alpha$; however, simulations by \citet{Esquivel2019,Debrecht2022} have suggested that ENAs might not fully explain observed Lyman-$\alpha$ transit signatures. {\rc It is important to note that different simulations have used distinct approaches to the inclusion of ENAs in the modelling. As discussed in Section~\ref{sec:future}, this results in discrepant conclusions about the importance (or not) of ENAs and this area remains under debate in the literature. } Finally, simulations have noted the importance in the ram-pressure of the stellar-wind in radially accelerating the escaping gas to sufficiently high-velocities to match the observed absorption seen at $\gtrsim 100$~km~s$^{-1}$ \citep{Bisikalo2013,McCann2019,Khodachenko2019,Debrecht2020,Carolan2021}. Thus, when one tries to match observations of specific systems, key input parameters for simulations such as the stellar EUV flux and wind properties are directly unobservable, and can only be inferred from modelling or empirical scalings. 

At this stage, what is not clear is how all these different processes compete to explain the properties of observed Lyman-$\alpha$ transits, the non-detections, and what properties of the atmospheric escape can actually be extracted from observations. In this work we aim to build on our knowledge gained from individual simulations to present a qualitative description of the physics of Lyman-$\alpha$ transits. Our aim is to characterise how ionization, stellar wind acceleration and charge exchange all interact within a phenomenological model in order to provide an understanding of (i) how these physical processes control the observability of a Lyman-$\alpha$ transit and (ii) what information about atmospheric escape a Lyman-$\alpha$ transit encodes. This understanding will allow more targeted use of 3D radiation-hydrodynamic simulations to compare to observations and an understanding of how to construct observational tests of atmospheric escape using Lyman-$\alpha$ transits.  

\section{Phenomenological Tail Model}
\begin{figure*}
\centering
\includegraphics[width=\textwidth, trim=0 0.7cm 0 0,clip]{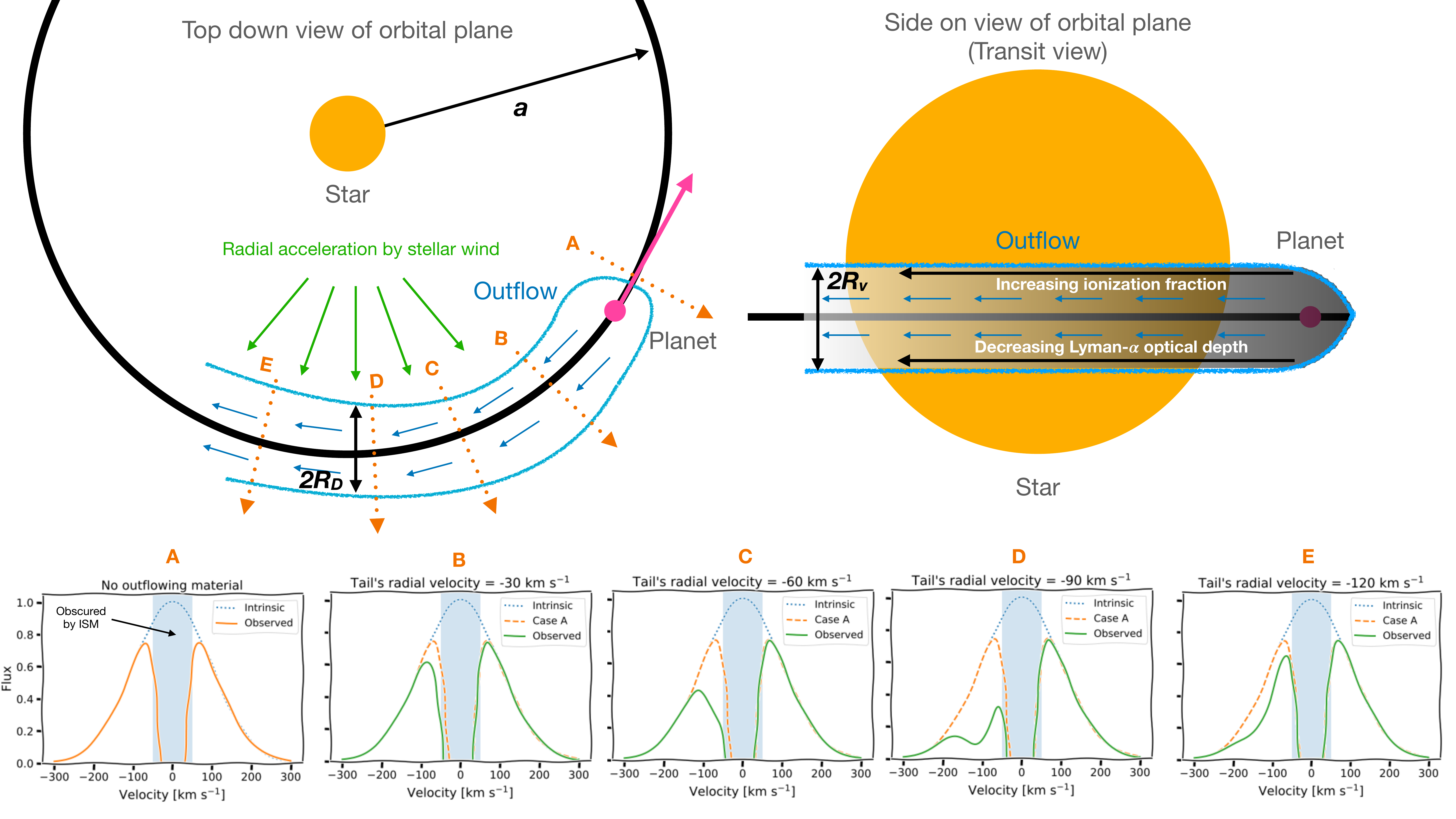}
\caption{A schematic diagram of the outflowing gas shaped into a planetary tail, and the origin of a Lyman-$\alpha$ transit. The top, left and right panel shows the system geometry when observed top-down (left), or side-on (right). The cloud of escaping hydrogen moves away from the planet with a bulk velocity ($u_t$ - represented by the blue arrows) and is progressively ionized, reducing the Lyman-$\alpha$ optical depth. Schematic Lyman-$\alpha$ transmission spectra are shown at different phases, labelled A--E, in the bottom panels. The blue dotted line shows the intrinsic stellar Lyman-$\alpha$ line, and the solid lines show the observed Lyman-$\alpha$ line, including absorption by the ISM (shaded region) and the planetary outflow. In panels B--E, the profile unobscured by the planetary outflow (A) is repeated for reference (orange dashed line). As the gas in the tail is radially accelerated up to $-90$km~s$^{-1}$ (B$\longrightarrow$D) we see the transit signature appear in the blue-wing of the Lyman-$\alpha$ line with an increasing transit depth. Finally, the transit depth drops in panel E as the tail has become too ionized and the Lyman-$\alpha$ optical depth has fallen. }\label{fig:tail_sketch}
\end{figure*}

Motivated by the observations and recent hydrodynamic simulations we consider the behaviour of an atmospheric outflow on scales greater than the planet's Hill radius, $R_{\rm Hill} \approx a(M_p/3M_*)^{1/3}$, where the planet has mass $M_p$ and it orbits a star of mass $M_*$ at semi-major axis $a$. Beyond the Hill radius, the planet's outflow is shaped into a cylindrical tail (with an elliptic cross-section) by tidal/rotational forces, ram pressure from the stellar wind and/or radiation pressure.
Since we are focused on close-in planets, where hydrodynamic atmospheric escape takes place, we know that the thermal velocities of the gas and planetary escape velocity, which are tens of km/s at most, are small compared to the planet's orbital velocity. This means that gas leaving the planet ends up on orbits around the star with only small angular momentum and energy differences relative to the planet. Thus, we approximate the outflow as occupying a narrow cylinder such that the range of semi-major axes of gas parcels  $\Delta a / a \ll 1$. This approximation is easily satisfied by the focus on close-in planets. In reality the tail should slowly spiral out, but as we discuss in Section~\ref{sec:future} this is generally a small deviation, and as such will be considered in future work. We further consider that atmospheric escape will inject mass into this cylinder at a rate of $\dot{M}_w$ with a velocity $u_t$. In order to unveil the underlying physics of a Lyman-$\alpha$ transit, we consider three effects (i) how neutral hydrogen injected into the tail ionizes, and if it becomes ionized, recombines; (ii) how the tail is radially accelerated towards an observer, and (iii) how charge exchange with the stellar wind produces fast ($\sim -100$~ km~s$^{-1}$) neutral hydrogen atoms. A sketch of the tail model is shown in Figure~\ref{fig:tail_sketch}. 

We take the cross-section of the tail to be an ellipse with the axis in the orbital plane of size $R_D$, and the axis perpendicular to the orbital plane of size $R_v$. We take the bulk velocity of the gas in the tail relative to the planet to be constant and the mass-flux through the tail to be constant with a value $\dot{M}_w$. This allows us to find the density in the tail as $\rho=\dot{M}_w/\pi u_t R_D R_v$, with $u_t$ the velocity of the material as it flows along the tail. To understand the basic physics we shall assume that the gas injected into the tail is composed exclusively of neutral, atomic hydrogen. This is because for typical {\rc sub-Neptune} systems the flow timescale to the planet's Hill sphere ($R_{\rm Hill}/u_t$, 1--10 hours for typical sub-Neptunes) is shorter than the typical timescale for ionization of a hydrogen atom (5--50 hours for typical sub-Neptunes around Sun-like stars). However, as we show later, when this approximation breaks (at extreme irradiation levels), it does not strongly change our results. {\rc We do note this approximation will overestimate our transit depths, and it is worst for those planets with deep gravitational wells where the time it takes for a fluid parcel to reach the Hill sphere can exceed many sound-crossing times due to sub-sonic launching.}.   In order to understand if this tail of gas will subsequently give rise to a detectable Lyman-$\alpha$ transit we must consider how the hydrogen gas is subsequently ionized in the stellar EUV field, whether it will be accelerated to sufficiently high velocities to be observed obscuring the blue-wing of the stellar Lyman-$\alpha$ line and whether charge exchange will produce a significant optical depth in stellar wind neutral hydrogen atoms.

\subsection{Tail Geometry}
We can estimate the geometry of the cylinder's elliptical cross-section by considering the dominant physical processes that sculpt its shape. Since we are considering the tail in a frame co-rotating with the planet any motion of the gas in the cylinder is going to be deflected by the Coriolis force. A parcel of gas outflowing from the planet at relative velocity $\vec{u_t}$ feels a Coriolis acceleration of $\ddot \vec{r_C} = -2\vec{\Omega} \times \vec{u_t}$, where $\vec{\Omega}$ is the Keplerian angular velocity at the planet's orbital separation ($a$) and $\Omega = |\vec{\Omega}|$.  The parcel is deflected onto an azimuthal orbit (i.e., into the cylinder) on a timescale of $t_C \sim u_t/|\ddot \vec{r_c}|$, so that taken as a whole, the gas reaches a maximum radial distance $\sim$ $u_t t_C \sim u_t/\Omega$.  Hence, 
\begin{equation}
    R_D = f \frac{u_t}{\Omega}
\end{equation}
where  $f$ is a constant of proportionality. The value of $f$ depends on the distribution of angular momenta, energies and pressure gradients of the launched gas. In the case of collisionless particles all launched with the same angular momenta but different energies, the epicyclic approximation can be used to show $f\approx 1/2$. Alternatively, we can consider the particles launched parallel to the direction of the planet's travel with a range of different angular momenta and determine $f$ from the Keplerian shear to find $f\approx 2/3$. Thus, these values of $f$ represent its plausible ranges.  As we show later, our results are actually independent of the exact value of $R_D$, and hence $f$. Thus for the remainder of this work we set $f=1/2$, which is in reasonable agreement with he simulations of \citet{McCann2019}. 
For the vertical extent of the tail $R_v$ we  consider the height a gas parcel fired from the top of the Hill sphere ($R_{\rm Hill}$) would reach at a velocity $u_t$ under the influence of stellar gravity. Performing this analysis yields: 
\begin{equation}
    R_v = \sqrt{R_{\rm Hill}^2+\left(\frac{u_t}{\Omega}\right)^2} \approx \frac{u_t}{\Omega} \label{eqn:height}
\end{equation}
where the last approximation holds for many planets. This approximation arises as $u_t$ is of order the escape velocity ($v_{\rm esc}$) from the planet's atmosphere for a thermally launched wind. Thus, we can see that the last approximation holds by recasting $R_{Hill}$ as $R_{Hill} \simeq \sqrt{R_p/R_{\rm Hill}}v_{\rm esc}/\Omega $, where $R_p$ is the planet's radius. Therefore, the first term in equation \ref{eqn:height} is either smaller or, at  maximum, comparable to the second term. This approximation shows that $R_D/R_v\approx 1/2$. We note that unlike $R_D$, the tail's height scale does play a role and will need to be calculated in more detail in quantitative models, as we have neglected the contribution from pressure. However, this approximate size scale agrees with what would be obtained in hydrostatic equilibrium (e.g. $c_s/\Omega$, with $c_s$ the isothermal sound speed) as $u_t\sim c_s$ since the planetary wind is thermally launched. 

\subsection{Ionization of the tail}
\label{sec:ionization}
In the photoevaporation model the wind is launched by the absorption of EUV photons close to the planet. Since it is this material, launched from close to the planet, that ultimately ends up in the tail, the tail is going to be optically thin to stellar EUV photons, otherwise a wind could not be launched with such high densities. We can confirm this with a simple calculation: the optical depth to 13.6eV photons (the ionizing energy of hydrogen) through the tail is approximately $\tau_{\rm EUV}\sim 2 n_{\rm HI}\sigma_{\rm 13.6\,eV} R_D$ where $n_{\rm HI}$ is the number density of neutral hydrogen and $\sigma_{\rm 13.6\, eV}$ is the cross-section to ionizing photons. Rewriting the number density in terms of the mass-flux we find:
\begin{equation}
\tau^{\rm photo}_{\rm EUV} \!\approx \!0.15 \mathcal{N}\! \left(\frac{\dot{M}_w}{10^{10}~{\rm g/s}}\right)\!\left(\frac{u_t}{\rm 10~km/s}\right)^{-2}\!\! \left(\frac{P}{10~{\rm days}}\right)^{-1}\!\! \left(\frac{R_D/R_v}{1/2}\right)
\end{equation}
with $\mathcal{N}$ the neutral fraction and $P$ the period.  Alternatively if one were to use a different mass-loss mechanism, such as the core-powered mass-loss model \citep[e.g.][]{Ginzburg2018}, where the flow is not launched by stellar EUV, we can still check the tail will be optically thin. Using standard parameters \citep[e.g.][]{Gupta2021}, we find:
\begin{equation}
\tau^{\rm CPML}_{\rm EUV} \!\approx \!0.38 \mathcal{N}\! \left(\frac{\dot{M}_w}{10^{9}~{\rm g/s}}\right)\!\left(\frac{u_t}{\rm 2~km/s}\right)^{-2}\!\! \left(\frac{P}{10~{\rm days}}\right)^{-1}\!\! \left(\frac{R_D/R_v}{1/2}\right)
\end{equation}

Thus even under the conservative assumptions of the maximal EUV absorption cross-section (remember $\sigma \propto \nu^{-3}$) and fully neutral hydrogen the tail is optically thin to stellar EUV photons for winds launched by either the photoevaporation or core-powered mass-loss model. So as gas progresses along the cylinder it will absorb stellar EUV photons with minimal self-shielding and be progressively ionized. In the optically thin limit, the evolution of the ionized fraction is given by:
\begin{equation}
    \frac{{\rm D}X}{{\rm D}t} = (1-X)\Gamma - nX^2\alpha_A \label{eqn:with_recomb}
\end{equation}
where $X=1-\mathcal{N}$ is ionization fraction, $\Gamma$ is the optically thin photoionization rate and $\alpha_A$ is the case-A recombination coefficient\footnote{Note since the tail is optically thin to EUV photons, it will be optically thin to ionizing photons produced via ground-state recombination. Thus, we make the ``optically-thin'' nebula approximation, rather than the more commonly-used, but incorrect in this case, on-the-spot approximation \citep[see, e.g.][]{Osterbrock2006}}. The optically thin photo-ionzation rate is given by:
\begin{equation}
    \Gamma = \int_{13.6\,{\rm eV}}^{\infty} \frac{L_{\nu}}{4\pi a^2 h \nu}\sigma_\nu\, {\rm d} \nu \approx \frac{L_{\nu}}{4\pi a^2 \overline{h\nu}}\sigma_{\overline{\nu}}
\end{equation}
where $\overline{h\nu}$ is a representative EUV photon which we take to be 20eV (a typical representative energy, e.g. \citealt{MurrayClay2009}), and $\sigma_{\overline{\nu}}$ is the photoionization cross-section for these photons. In using the case-A recombination coefficient we have assumed that all ionizing recombination photons can freely escape the tail. As we have demonstrated the tail is optically thin to 13.6~eV photons, this approximation is justified.  As we shall see when we numerically solve Equation~\ref{eqn:with_recomb}, we can typically drop the recombination term and it is convenient to re-arrange for the neutral fraction. Thus, the governing equation for the steady-state neutral fraction along the cylinder approximately becomes:
\begin{equation}
    u_t\frac{\partial \mathcal{N}}{\partial \ell} = -\mathcal{N}\Gamma \label{eqn:neutral}
\end{equation}
where $\ell$ is the distance along the cylinder's axis. Equation~\ref{eqn:neutral} can be solved to give the neutral fraction along the cylinder as:
\begin{equation}
    \mathcal{N}= \mathcal{N}_0\exp\left(-\frac{\Gamma}{u_t}\ell\right)
\end{equation}
with $\mathcal{N}_0$ the neutral fraction of the gas the planet injects into the cylinder.  A Lyman-$\alpha$ transit will be present if the tail is optically thick to Lyman-$\alpha$ photons. Thus as the gas in the tail recedes from the planet and becomes progressively ionized and hence more optically thin to Lyman-$\alpha$ photons, the tail will become transparent and cease to give rise to a transit signature. We can calculate the length along the tail at which this will happen by finding the distance relative to the planet a fluid parcel has travelled before it becomes optically thin to Lyman-$\alpha$ photons. Namely, finding the value of $\ell$ when $2R_Dn_{\rm HI}\sigma_{Ly\alpha}\sim 1$. Solving for this length we find:
\begin{equation}
    \ell_{\tau_{\alpha}=1} = \frac{u_t}{\Gamma}\log\left(\frac{2\sigma_{Ly\alpha}\dot{M}_w\mathcal{N}_0}{\pi R_v \mu u_t}\right) = \ell_\Gamma \log {\mathcal{T}}_p
    \label{eqn:lya_length}
\end{equation}
where $\ell_\Gamma$ encodes the mean-free path of a neutral hydrogen atom before it's photoionized and $\mathcal{T}_p$ includes all the terms in the $\log$ term. This term represents the Lyman-$\alpha$ optical depth through the cylinder at its start - i.e. where the planet's Hill sphere connects to the cylinder.  Now if we consider the basic outcome of a transit we determine two quantities: the transit depth and the transit duration. The transit depth  is approximately\footnote{Technically, we should evaluate $\ell$ over the chord of the planet's orbit; however, since $R_*/a\ll 1$ this is an unimportant correction.}:
\begin{equation}
{\rm Depth} \approx \frac{2R_v\min\left(R_*,\ell_{\tau_{\alpha}=1}\right)}{\pi R_*^2}
\end{equation}
and the transit duration is approximately
\begin{equation}
    {\rm Duration} \approx \frac{R_* + \ell_{\tau_{\alpha}=1}}{2\pi a} P.
\end{equation}
As the cylinder's height ($R_v$) increases as the planet's separation increases, we arrive at what appears initially to be a quite counter-intuitive result: weaker irradiation levels, which drive weaker mass-loss yield deeper, longer Lyman-$\alpha$ transits. Thus, the strength of a Lyman-$\alpha$ transit is actually {\it anti-correlated} with the magnitude of escape. However, with our model we can understand this result quite easily. Even before we get onto the complication of obscuration of the Lyman-$\alpha$ line core and transits in the blue-wing of the line it is clear that the primary physical parameter a Lyman-$\alpha$ transit actually measures is not the mass-loss rate, but rather the length along the cylinder that a gas particle travels before it is photo-ionized ($\ell_\Gamma$). The importance of ionization was noted in the 3D particle simulations of \citet{Bourrier2013} and \citet{Bourrier2016}, where they found higher escape rates were required to counter-balance higher photoionization rates to give similar absorption depths. We further see that the Lyman-$\alpha$ tail-length is only logarthmically sensitive to the mass-loss rate, meaning that Lyman-$\alpha$ transits provide weak observational constraints on the mass-loss rates. Or, only with a very accurate model of an outflow and its interaction with the circumstellar environment could the mass-loss rate be reliably estimated from Lyman-$\alpha$ transits (although see our discussion in Section~\ref{sec:discuss}, on how one might be able to actually constrain mass-loss rates using velocity resolved transits). 

Does this insight mean Lyman-$\alpha$ transits are useless observational probes of atmospheric escape? Absolutely not! The fact that the transit duration is primarily sensitive to $\ell_{\tau_{\alpha}=1}$ means it provides a clean way of measuring the bulk velocity of the gas the planet injects into the tail ($u_t$). {\rc The accuracy to which this velocity can be observationally determined will ultimately be limited by the accuracy to which the photoionization rate ($\Gamma$) can be measured, something we discuss in Section~\ref{sec:discuss}.}

\subsection{Radial acceleration of the tail}
We now must turn our attention to one of the more curious aspects of observed Lyman-$\alpha$ transits. Primarily due to interstellar absorption, the core of the Lyman-$\alpha$ line is usually inaccessible out to $\sim \pm50$km~s$^{-1}$. Since the typical temperatures of the outflow are expected to be in the range $10^4$~K, one expects the absorption in the tail to be Doppler broadened by only a few 10~km~s$^{-1}$. The fact we see absorption at velocities around $-100$km~s$^{-1}$ in the blue wing implies the tail is either accelerated away from the star to these velocities\footnote{The fact the absorption is not symmetric in both the blue- and red-wing means Lyman-$\alpha$ transits do not primarily arise due to low-velocity gas absorbing in optically thick line-wings.} or the absorption is arising from charge exchange between the tail and stellar wind protons \citep[e.g.][]{Holmstrom2008,Tremblin2013}. Recent simulations \citep[e.g.][]{Khodachenko2017,Esquivel2019} suggest that for short-period planets orbiting Sun-like stars, energetic neutral atoms arising from charge exchange are not significant enough to explain the observed Lyman-$\alpha$ absorption, implying acceleration of the tail is the most likely scenario. Two possible physical mechanisms have been identified: acceleration by ram pressure from the stellar wind and acceleration by radiation pressure in the Lyman-$\alpha$ line \citep[see review by][]{Owen2019}.  The momentum flux incident on the tail due to the stellar wind is $\dot{M}_*u_*/4\pi a^2$ (with $\dot{M}_*$ and $u_*$ the stellar wind mass-loss rate and velocity respectively) and the momentum flux due to Lyman-$\alpha$ photons is $L_{Ly\alpha}/4\pi a^2 c$ (with $L_{Ly\alpha}$ the stellar luminosity in the Lyman-$\alpha$ line and $c$ the speed-of light). Thus the ratio of stellar wind to Lyman-$\alpha$ momentum flux is:
\begin{equation}
    \left(\frac{\dot{M}_*}{L_{Ly\alpha}}\right)u_*c \approx 30 \left(\frac{\dot{M}_*}{\dot{M}_\odot}\right)\left(\frac{L_{Ly\alpha}}{2\times10^{28}~{\rm erg~s^{-1}}}\right)^{-1}\left(\frac{u_*}{150~{\rm km~s^{-1}}}\right)
\end{equation}
where we have evaluated this for typical solar parameters at the orbital separation for a typical close-in exoplanet\footnote{As the stellar wind velocity increases with distance.}. Thus, we see that for Sun-like stars we would expect stellar wind acceleration to dominate over radiation pressure, and this conclusion is in agreement with simulations \citep[e.g.][]{Khodachenko2019,Debrecht2020,Carolan2021}. Thus, in this work, for simplicity we stick to accelerating the tail through the ram pressure interaction with the stellar wind. Because the outflow is generally optically thick to line-center Lyman-$\alpha$ photons over the observable tail, radiation pressure typically acts as a surface force, and our framework can be adapted to include radiation pressure by replacing the force in Equation \ref{eqn:radforce}.  We discuss charge-exchange within our framework in Section \ref{sec:enas}.    The radial force per unit-length ($f$) that the stellar wind applies to the cylinder as a ram-pressure is:
\begin{equation}
    f = 2\rho_*(u_*-u_r)^2R_v
\end{equation}\label{eqn:radforce}
where $u_r$ is the radial velocity of the cylinder and $\rho_*$ is the density of the stellar wind at the location of the planet. This allows us to write the radial acceleration on a fluid parcel in the cylinder, in the inertial frame, as (note that as we are assuming that the tail's distance from the planet does not change significantly, we are also implicitly assuming stellar gravity and the centrifugal force balance):
\begin{equation}
\frac{{\rm d}u_r}{{\rm d}t} = \frac{2\rho_* R_v u_t}{\dot{M}_w}\left(u_*-u_r\right)^2
\end{equation}
noting that $u_t\equiv{\rm d}\ell/{\rm d}t$ allows us to write an expression for how the radial velocity varies as a function of length along the cylinder:
\begin{equation}
    \frac{{\rm d}u_r}{{\rm d}\ell} = \frac{2\rho_*R_v}{\dot {M}_w}\left(u_*-u_r\right)^2
\end{equation}
whence solved, with $u_r=0$ at $\ell=0$, yields:
\begin{equation}
    u_r= \frac{2\rho_*R_v u_* \ell /\dot{M}_w}{2\rho_*R_v  \ell /\dot{M}_w + u_*^{-1}} \label{eqn:ur}
\end{equation}
It is instructive to combine our results from the previous section and introduce the concept of the ionization length ($\ell_\Gamma$) into Equation~\ref{eqn:ur}. Doing this we find:
\begin{equation}
    \frac{u_r}{u_*} = \frac{ \left(\ell/\ell_{\Gamma}\right)}{ \left(\ell/\ell_{\Gamma}\right) + S_p} \label{eqn:ur_ustar}
\end{equation}
where we have introduced a dimensionless variable:
\begin{equation}
    S_p \equiv \dot{M}_w/(2\rho_*R_v\ell_{\Gamma}u_*) \label{eqn:Sp}
\end{equation} which characterises the strength of the planetary wind compared to the stellar wind: it is the ratio of the planetary mass-loss to the stellar wind mass-loss that passes through the cylinder over one ionization-length. We can understand the combined implications of  Equation~\ref{eqn:lya_length} and Equation~\ref{eqn:ur_ustar} in terms of the observability of a Lyman-$\alpha$ transit in the blue-wing. To observe a Lyman-$\alpha$ transit in the blue-wing the radial velocity of gas in the cylinder must have been accelerated to a sufficiently high-velocity such that it is outside the line core ($u_r \sim u_*$), but by the time that's happened the gas in the cylinder must still be opaque to stellar Lyman-$\alpha$ photons, or $\ell < \ell_{\tau_\alpha=1}$. 
\begin{figure}
    \centering
    \includegraphics[width=\columnwidth]{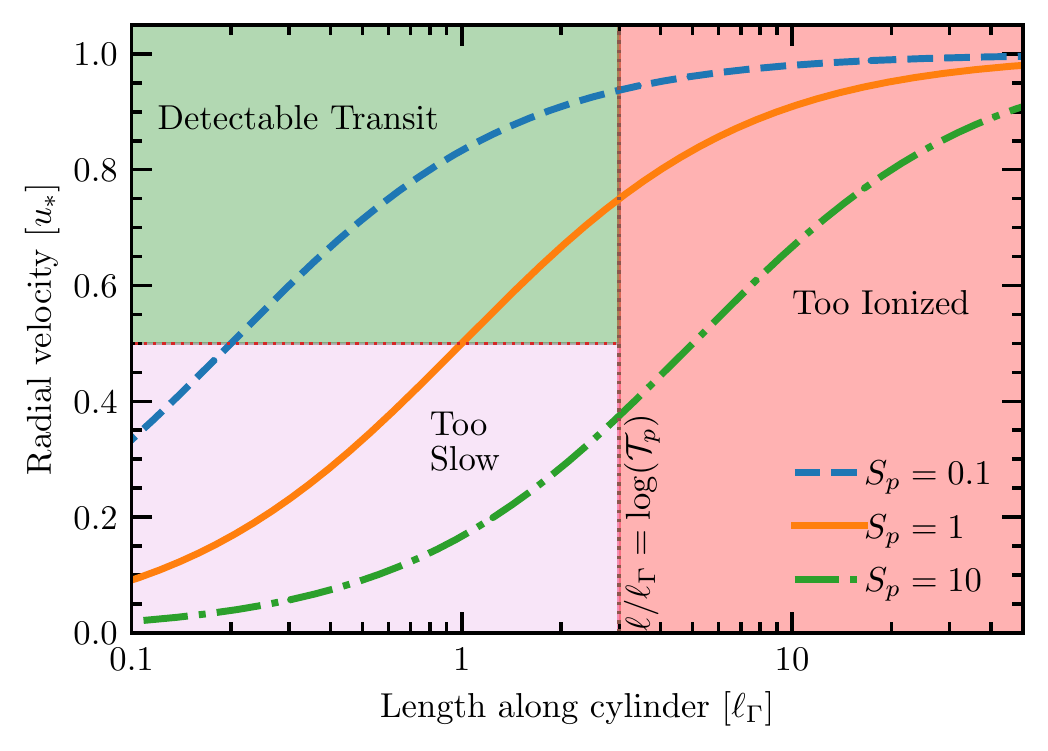}
    \caption{The lines show the evolution of the cylinder's radial velocity (plotted in units of the stellar wind velocity) as a function of length along the cylinder (plotted as a function of the ``ionization length'' introduced in Equation~\ref{eqn:lya_length}) for different values of the strength of the planetary wind compared to the stellar wind ($S_p$ -- Equation~\ref{eqn:Sp}). A detectable transit (as indicated by the green region) requires the cylinder obtains a radial velocity that is a significant fraction of the stellar wind velocity (taken here to be a half, purely for a demonstration) before it becomes optically thin to Lyman-$\alpha$ photons (i.e. when $\ell/\ell_\Gamma > \log\mathcal{T}_p$). The red region indicates where the cylinder is too ionized (and therefore transparent to Lyman-$\alpha$ photons) and would not yield any obscuration of the star whereas the magenta region shows points in the cylinder which are opaque to Lyman-$\alpha$, but do not have sufficient radial velocity to be observed in the blue-wing, although a transit would be observable in a high-RV system (e.g. $\gtrsim 50$~km~s$^{-1}$), where the line-core is observable.     }
    \label{fig:dimensionless_plot}
\end{figure}
We can understand this graphically in Figure~\ref{fig:dimensionless_plot}, where a detectable transit (green region) requires the gas in the cylinder to achieve a significant radial velocity (e.g. an appreciable fraction of the stellar wind velocity) before it is so ionized that it becomes transparent to Lyman-$\alpha$ photons. Of course this neglects the possibility that the cylinder is so optically thick that the Lorentzian-wings provide an appreciable optical depth at high velocity. However, this model underlines the basic picture of how we observe Lyman-$\alpha$ transits, and explains that the non-detection of a Lyman-$\alpha$ transit does not necessarily mean that the planet is not undergoing atmospheric escape, or doesn't possess a hydrogen dominated atmosphere. In fact, as we show later (in Section~\ref{sec:discuss}), we can reproduce the observed non-detections for sub-Neptunes, simply due to the fact that the escaping gas is photoionized before the stellar wind can accelerate it.  Finally, we note that while our results are sensitive to the assumed height of the cylinder ($R_v$) they are independent of the depth of the cylinder. This result is expected as both the optical depth and radial acceleration only depend the mass per unit area along the cylinder, not how that mass is distributed radially. 

\begin{figure*}
\centering
\includegraphics[width=0.99\textwidth]{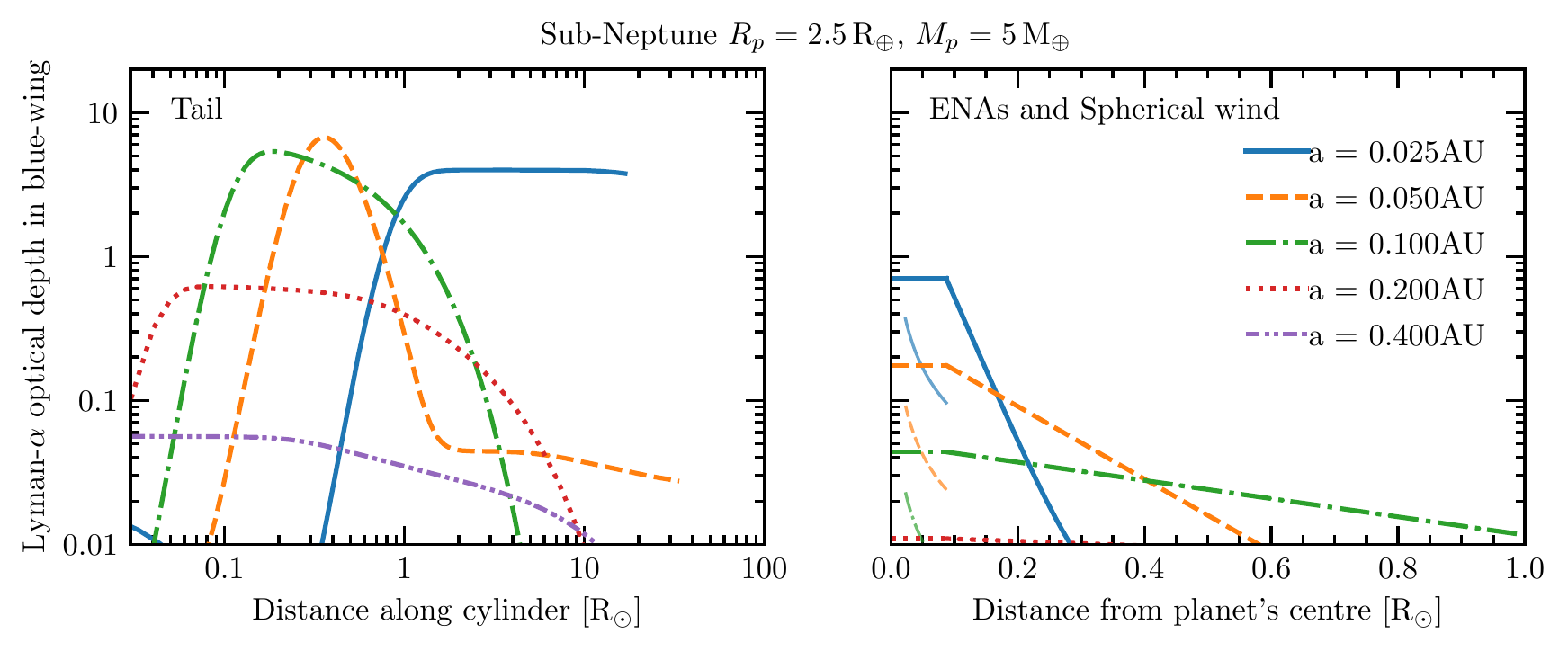}
\includegraphics[width=0.99\textwidth]{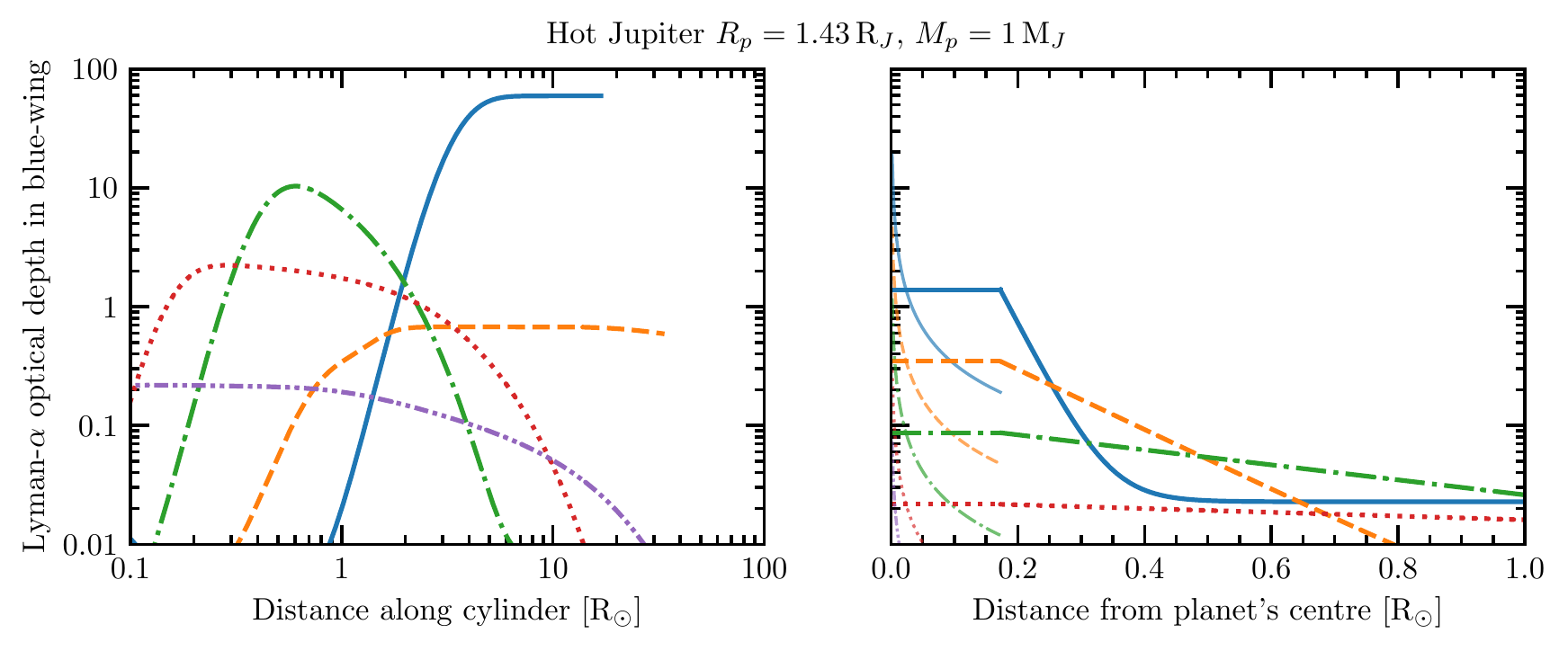}
\caption{The optical depth in the blue-wing of the Lyman-$\alpha$ line (taken to be averaged between -50 and -150~km~s$^{-1}$) as a function of distance due to planetary material in the tail (shown in the left panels) and the right-panel shows it arising due to ENAs (thick) and the spherical planetary wind residing inside the planet's Hill sphere (thin)  for planets residing at different orbital locations around a Sun-like star. The top panels shows results for a typical sub-Neptune and the bottom panels results for a Jupiter.  For the planets at the shortest orbital periods the line reaches a constant, high optical depth, due to the tail reaching recombination equilibrium as described in the text. Note the different scales on the x-axes between the left and right panels---ENAs and the spherical wind (right) are important on smaller radial scales than the tail (left).}\label{fig:optical_depth}
\end{figure*}

\subsection{Additional sources of Lyman-$\alpha$ optical depth}
The interaction between the planetary outflow and the stellar wind not only results in an acceleration, but also in charge exchange between stellar wind protons and neutral hydrogen in the outflow. This process generates neutral hydrogen atoms that originated in the stellar wind and thus have a high velocity. These energetic neutral atoms (ENAs) have been proposed as the origin of the $\sim - 100~\mathrm{km~s^{-1}}$ blue-shifted Lyman-$\alpha$ absorption \citep[e.g.][]{Holmstrom2008,Kislyakova2014,Khodachenko2019} in certain planets. However, the results from hydrodynamic simulations suggests that they cannot explain all of the observed features \citep[e.g.][]{Khodachenko2017}. Below we explore these results. Specifically, we find ENAs are likely to play an important, if not a dominant, role in the early stages of Lyman-$\alpha$ transits, particularly with the generation  of high velocity neutrals when the stellar wind interacts with gas close to the planet's Hill sphere. However, once the stellar wind begins to radially accelerate the tail to tens of km~s$^{-1}$ ENAs become sub-dominant compared to absorption in the tail. 

\subsubsection{Contribution from ENAs to the Lyman-$\alpha$ optical depth}\label{sec:enas}

{\rc In the following, we adopt a fluid description to the production of ENAs. As discussed in Section~\ref{sec:future} this approach gives smaller ENA Lyman-$\alpha$ optical depths compared to other approaches such as multi-fluid approaches \citep[e.g.][]{Khodachenko2019} or particle approaches \citep[e.g.][]{Lavie2017,BenJaffel2022}. Since all approaches used in simulations approximate some part of the problem it's unclear which approach is more accurate; however, it is worth noting the model adopted here may underestimate the ENA Lyman-$\alpha$ optical depth}. 

The rate of production of ENAs per unit volume is given by ($\Gamma_{\rm ENA}$):

\begin{equation}
    \Gamma_{\rm ENA} = \beta\left(n_{+,*}n_{0,p}-n_{0,*}n_{+,p}\right)
\end{equation}
where $\beta$ refers to the reaction rate and $n$ refers to the number density of different components of the stellar/planetary wind with $_+$, $_0$ subscripts referring to ionized and neutral hydrogen, and $_*$, $_p$ subscripts referring to stellar and planetary origin, respectively. These densities are those in the vicinity of the mixing layer, and for the star are approximately the value in the stellar wind at the planet's location, and for the planet are approximately the densities in the tail. \citet{Tremblin2013} calculated $\beta = 4\times10^{-8}$ cm$^3$~s$^{-1}$ based on a stellar wind with a temperature of $10^6$~K and planetary wind temperature of $7000$~K. 

In the region where the stellar and planetary wind are interacting, charge exchange will drive the stellar wind protons towards a collisional equilibrium on a time-scale $t_{\rm col}$:

\begin{eqnarray}
t_{\rm col}    \sim &&\!\!\!\!\!\! \frac{1}{\beta n_{0,p}} = \frac{\pi u_t^3 \mu}{2\beta\Omega^2\mathcal{N}\dot{M}_w}\nonumber \\
\sim&& \!\!\!\!\!\! 100~{\rm s}\, \mathcal{N}^{-1} \left(\frac{u_t}{10~{\rm km~s^{-1}}}\right)^3 \left(\frac{P}{10~{\rm days}}\right)^2\left(\frac{\dot{M}_w}{10^{10}~{\rm g~s^{-1}}}\right)^{-1}
\end{eqnarray}
As pointed out by \citet{Tremblin2013}, this timescale to reach collisional equilibrium is very short compared to both the flowscale $\sim R_v/u_t$ which, using Equation~\ref{eqn:height}, is of order the orbital period and transit duration. Therefore, in the mixing region between the planetary and stellar wind we can assume collisional equilibrium. Taking the stellar wind density to be much smaller than the planetary wind density, charge exchange with stellar wind protons does not change the ionization fraction of the planetary wind. This means that the ionization fraction of the stellar wind hydrogen quickly reaches the same value as the ionization fraction in the planetary wind. This allows us to write the density of ENAs in the mixing region as:
\begin{equation}
    n_{\rm ENA} \approx \mathcal{N} n_*
\end{equation}

Strictly speaking $n_*$ in the above equation is the post-shocked stellar wind density. However, at this distance of most observed close-in exoplanets, the stellar-wind is only marginally super-sonic, {\rc such that the density and velocity} in the post-shocked region is similar to the stellar wind density itself. Now, the contribution of ENAs to the Lyman-$\alpha$ optical depth depends on the depth of the mixing region. The size of the mixing region has been studied analytically \citep[e.g.][]{Dyson1975,Raga1995}, for a spherical outflow interacting with a plane parallel flow (similar to our problem here), \citet{Raga1995} found the size of the mixing region to be a few-to-tens percent of the radius of curvature of the bow-shock $R_c$. \citet{Tremblin2013} performed simulations and measured the size of the mixing layer to be $\sim 0.1 \sqrt{n_p/n_*} \times$ the planet's radius. The interaction between the planetary and stellar wind cannot know about the planet's radius; we note that since \citet{Tremblin2013}'s simulations were performed in dimensionless units scaled to the planet's radius, we should translate their measurement into a fraction of the radius of curvature of the shock. The dependence of the mixing layer on the ratio of densities was attributed to the growth timescales of Kelvin-Helmholtz instabilities observed in their simulations. Thus, we write the depth of our mixing layer as $ D_{\rm mix} \sim 0.015\sqrt{n_p/n_*}R_c$, resulting in a Lyman-$\alpha$ optical-depth due to ENAs of:
\begin{equation}
    \tau_{\rm ENA} \approx 0.02 \mathcal{N} \sqrt{n_p n_*}\sigma_{Ly\alpha} R_v.
\end{equation}

This means that the ratio of the optical depth of ENAs to material in the tail is given by:
\begin{equation}
    \frac{\tau_{\rm ENA}}{\tau_{\rm tail}} \sim 0.04 \sqrt{\frac{n_*}{n_p}} \frac{\sigma_{Ly\alpha}(v_{\rm ENA})}{\sigma_{Ly\alpha}(u_r)}.
\end{equation}
Setting $v_{\rm ENA}$ to 100~km~s$^{-1}$, {\rc a crude estimate based on the fact collisions with planetary material will decelerate stellar material}, we find that for typical parameters, if $u_r\lesssim 25$~km~s$^{-1}$, ENAs dominate the optical depth over material in the tail; but for $u_r\gtrsim 25$~km~s$^{-1}$ planetary material that has been radially accelerated by the stellar wind dominates the Lyman-$\alpha$ optical depth. Thus, ENAs can provide an important contribution to the Lyman-$\alpha$ transits early, but the planetary material will dominate later on provided it can remain sufficiently neutral. 

\subsubsection{Contribution from inside the planet's Hill sphere}
Similar to ENAs early in the planetary transit, an additional contribution can arise: even though the material inside the planet's Hill sphere is unlikely to be strongly effected by the stellar wind and tidal forces, it can still be optically thick in the blue-wing \citep[e.g.][]{McCann2019}. This arises, simply because material in the Hill sphere is so optically thick in the line-core, that the Lorentzian-wings are also optically thick. We can estimate this contribution by considering the density profile to follow a $n\propto r^{-2}$ profile, appropriate for steady spherical flow with a constant velocity, which well approximates the outflow outside the sonic point (and will underestimate the optical depth interior to the sonic point). Thus, the optical depth as a function of distance from the planet ($\ell_p$) is given by:
\begin{equation}
    \tau = \int_{-\infty}^\infty\! \sigma_{\rm Ly\alpha}n_{\rm surf} \left(\frac{R_p}{\sqrt{x^2+\ell_p^2}}\right)^{2}{\rm d}x = \frac{\pi \sigma_{\rm Ly\alpha}n_{\rm surf}R_p^2}{\ell_p},
\end{equation}
where $n_{\rm surf}$ is the number density of neutral hydrogen at the planet's radius and $x$ is a co-ordinate along the line-of-sight to the star. To account for the fact the velocity in the sphere contains material gas travelling both away from and towards the observer {\rc we include this broadening in} the Lyman-$\alpha$ cross-section with an artificial thermal velocity of $\sqrt{2}u_t$. Finally, we compute the standoff distance of the shock between the stellar and planetary wind, if we determine that the planetary wind penetrates inside the Hill radius we assume that this region only contributes out to a radius of the standoff distance.  Clearly, very close to the planet this contribution will dominate over ENAs, but the area of the star blocked will be small. Thus, we need to explore realistic planetary types to understand whether ENAs or the material in the planet's vicinity dominates the initial stages of a Lyman-$\alpha$ transit. 

\subsection{Quantitative examples}
So far we have tried to work without specification of real parameters to illuminate the {\rc basic} physics at work. However, here we consider several real examples to give a sense of the scales. We will work exclusively within the photoevaporation model, considering a hot Jupiter like system ($R_p=1.43~$R$_J$, $M_p=1$~M$_J$) and a sub-Neptune like system ($R_p=2.5~$R$_\oplus$, $M_p=5$~M$_\oplus$), around a solar-type star which outputs $1\times10^{28}$~erg~s$^{-1}$ in the EUV and has a wind mass-loss rate comparable to the Sun's of $\dot{M}_\odot=2\times10^{-14}$~M$_\odot$~yr$^{-1}$ and a velocity of $150$~km~s$^{-1}$ at the location of the planet. The star-planet system is assumed to have no relative radial velocity to Earth. We assume the gas launched by the planet into the tail is pure atomic hydrogen, with a temperature of $10^{4}$~K and velocity of 10~km~s$^{-1}$. We estimate the planetary mass-loss rates from the energy-limited model:
\begin{equation}
    \dot{M}_w = \eta\frac{\pi R_p^3 F_{\rm EUV}}{GM_p}
\end{equation}
where we pick a constant efficiency of $\eta=0.1$. We include recombination in our solution for the ionization fraction (e.g. we numerically solve Equation~\ref{eqn:with_recomb} along the tail using the {\sc lsoda} ODE library, \citealt{lsoda})  to demonstrate that it's only important in a few special cases of extremely highly irradiated planets.

\subsubsection{Lyman-$\alpha$ optical depths}
In Figure~\ref{fig:optical_depth}, we show the average optical depths in the Lyman-$\alpha$ blue-wing for a sub-Neptune and hot Jupiter in orbits ranging from 0.025 to 0.4~AU around a Sun-like star. The transit is taken to occur along the star's equator, with an impact parameter of $b=0$. The left panels show the optical depths arising directly from planetary material in the tail, while the right panels show the contribution from ENAs and the material in a quasi-spherical wind emanating from the planet within its Hill sphere. These results show that long Lyman-$\alpha$ transits are associated with fairly weak irradiation levels, e.g. EUV fluxes around $\sim$500 erg~s$^{-1}$~cm$^{-2}$, at these fluxes the ionization rate of neutral hydrogen is slow enough to allow it to remain sufficiently neutral to 10-100's of planetary radii, yet are still able to be accelerated efficiently by the stellar wind. At very low-fluxes $\lesssim 30$ erg~s$^{-1}$~cm$^{-2}$ ($a \gtrsim 0.35$~AU for an old Sun-like star) the planetary winds are just too rarefied to present any detectable transit, while at high fluxes $\gtrsim 1000$~erg~s$^{-1}$~cm$^{-2}$ ($a \lesssim 0.06$~AU for an old Sun-like star) the tail is too rapidly ionized by the star before it can be accelerated into the blue-wing of the Lyman-$\alpha$ line. Generally, we also see that any transit will be primarily due to material accelerated towards the observer by the stellar wind in the tail; however, ENAs and the material close to the planet do add a non-negligible contribution early in the planet's transit, especially for the hot Jupiters. 

As expected, although we generally find recombination is unimportant (the approximately flat line at large distances for the 0.05 AU sub-Neptune is controlled by recombination), it does play an important role in strongly irradiated planets, particularly the hot Jupiters. This is because, while the ionization rates are higher, the densities are also higher due to the higher mass-loss rates (note recombination is proportional to $n^2$), thus the tail reaches ionization equilibrium quicker, producing a non-negligible Lyman-$\alpha$ optical depth. If we consider both the 0.025 AU sub-Neptune and hot Jupiter, we see that the planetary Hill sphere and ENAs could give rise to a Lyman-$\alpha$ signature during primary transit of the planet. Then at a distance of several stellar radii (and thus $\sim 10$ hours after primary transit) the tail is accelerated to a sufficiently high velocity towards the observer that it's optically thick in the blue-wing. Material this far from the planet is in ionization-recombination equilibrium. Thus, one would see the slightly unusual signature of a small Lyman-$\alpha$ transit concurrent with the planetary transit and then a second Lyman-$\alpha$ transit several hours later. In fact, the 3D radiation-hydrodynamic simulations of hot Jupiters in \citet{McCann2019} predicted such a feature, arising from identical physics - a primary Lyman-$\alpha$ transit from the optically thick tail of material in the Hill sphere and a delayed transit from material in the tail that's in recombination equilibrium but not accelerated into the blue-wing by the stellar wind until $\sim 10$ hours after primary transit. It is important to note that in our model its construction would produce a tail in ionization-recombination equilibrium that would wrap all the way around the planet's orbit. Ultimately, ram-pressure stripping, shear instabilities or some other process is likely to erode the tail before this can happen.  

\begin{figure}
    \centering
    \includegraphics[width=\columnwidth]{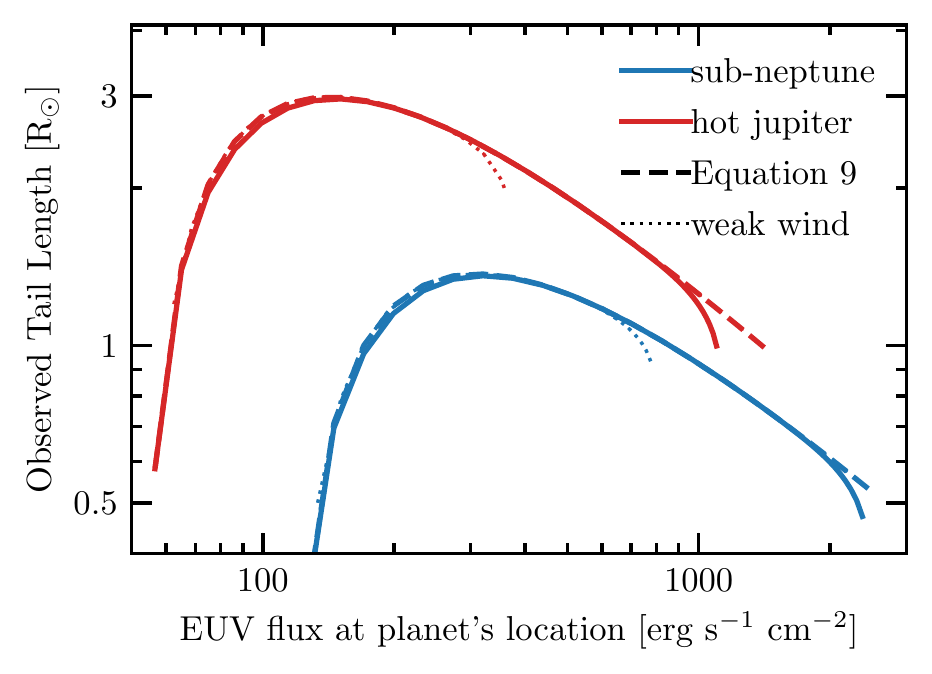}
    \caption{The length along the cylinder at which its optical depth to the observer becomes less than unity in the blue-wing for our typical sub-Neptune and hot Jupiter. This length is measured by averaging the Lyman-$\alpha$ optical depth between $-50$~km~s$^{-1}$ and $-150$~km~s$^{-1}$ and then finding the distance along the cylinder at which this average optical depth drops below unity.  The solid line shows a case for a solar stellar wind loss rate, while the dotted line shows the case where the wind is a factor of four weaker. At high-fluxes the stellar wind is unable to accelerate the tail to sufficiently high velocities before it becomes too ionized and the transit is no-longer detectable. The dashed line represents the theoretical model in Equation~\ref{eqn:lya_length} which neglects the role of the stellar wind, and any assumption about the velocity at which the Lyman-$\alpha$ line is observed and as such provides an upper limit to the tail length. }
    \label{fig:tail_length}
\end{figure}

Finally, we can use our models to assess how big of a correction the stellar wind makes to the observed Lyman-$\alpha$ tail lengths, when compared to our simple estimate in Section~\ref{sec:ionization} (Equation~\ref{eqn:lya_length}). Using our models from Figure~\ref{fig:optical_depth} we can compute the distance along the cylinder where the optical depth drops below unity and compare the result to the estimate in Equation~\ref{eqn:lya_length}. We ignore recombination in this calculation as it would either predict a gap between primary transit and a delayed obscuration or imply the tail fully wraps around the planet's orbit. The result is shown in Figure~\ref{fig:tail_length}, for both our sub-Neptune and hot Jupiter model as the solid lines. 
To compare to our model in Equation~\ref{eqn:lya_length}, to those where we account for the tail's radial acceleration and blue-shift, we have averaged the Lyman-$\alpha$ cross-section in Equation~\ref{eqn:lya_length} over 100 km~s$^{-1}$ about the line-core. This result is shown as the dashed line in Figure~\ref{fig:tail_length}. We find Equation~\ref{eqn:lya_length} provides a good agreement; however, at high-fluxes the simple theory model over predicts the length. This is easy to understand as it is a demonstration of our previous discussion: the tail rapidly disappears with increasing EUV flux as it is photoionized before it is accelerated into the blue-wing by the stellar wind. Finally, we also indicate the critical role of the stellar-wind in accelerating the tail into the blue-wing. The importance of this acceleration is demonstrated where we show the observed tail length for a model that has a stellar-wind mass-loss rate a factor of four lower than our nominal solar choice. In this case the range of EUV flux over which a Lyman-$\alpha$ tail could be observed is reduced by a factor of $\sim 4$ at high-fluxes. Thus, one could imagine scenarios where the stellar wind is too weak to ever observe a Lyman-$\alpha$ transit because photoionization controls when the transit ends (flow ionized into transparency), whereas the stellar wind controls when it begins (flow accelerated into the Lyman-$\alpha$ blue wing). The difference translates into the observed  tail length. By delaying the start of the transit, weaker stellar winds yield a shorter observed tail length. If the stellar wind is too weak for a detectable transit to begin before photoionization ends it, then a non-detection results.  

\subsubsection{Simple Light Curves}

\begin{figure}
    \centering
    \includegraphics[width=\columnwidth]{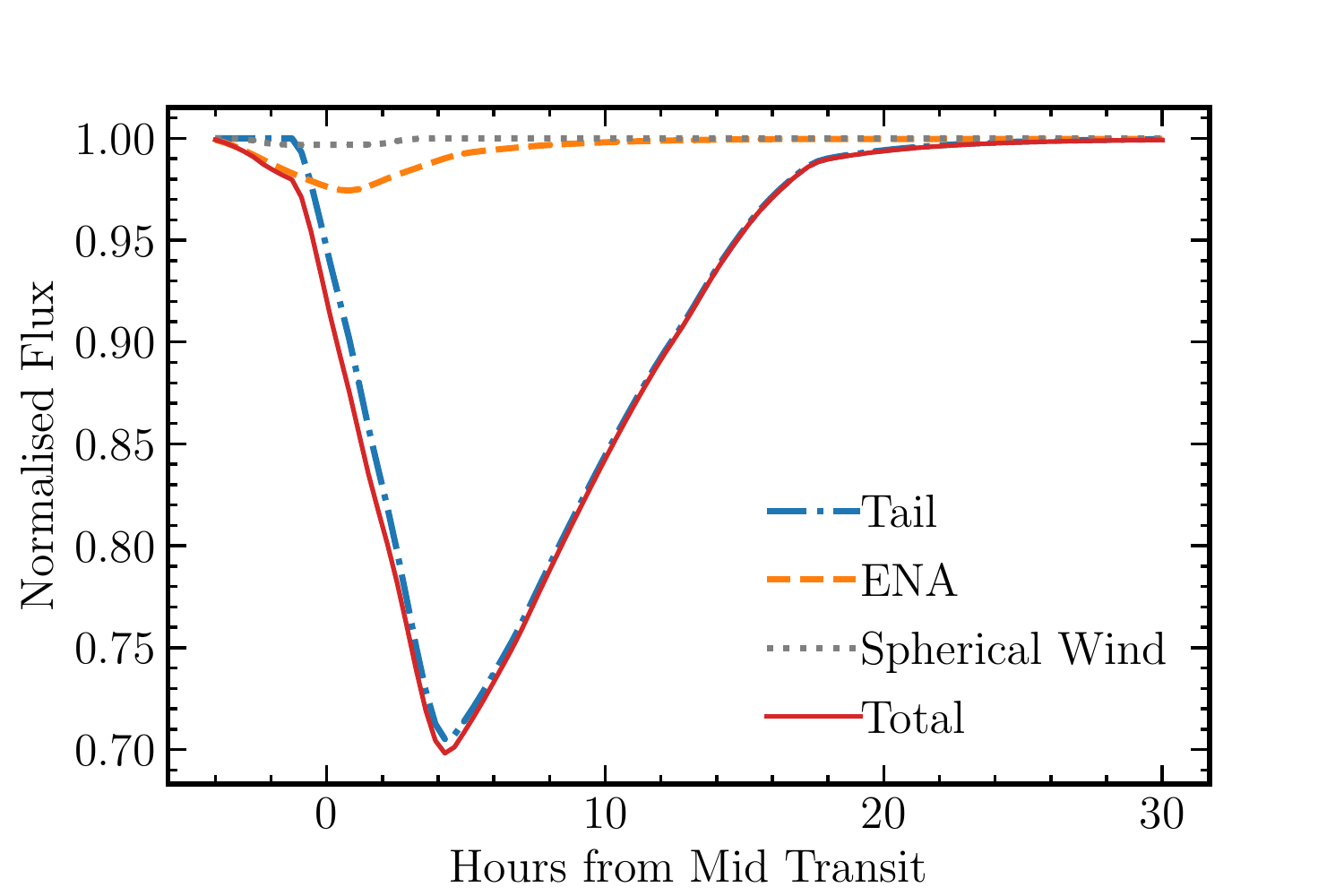}
    \includegraphics[width=\columnwidth]{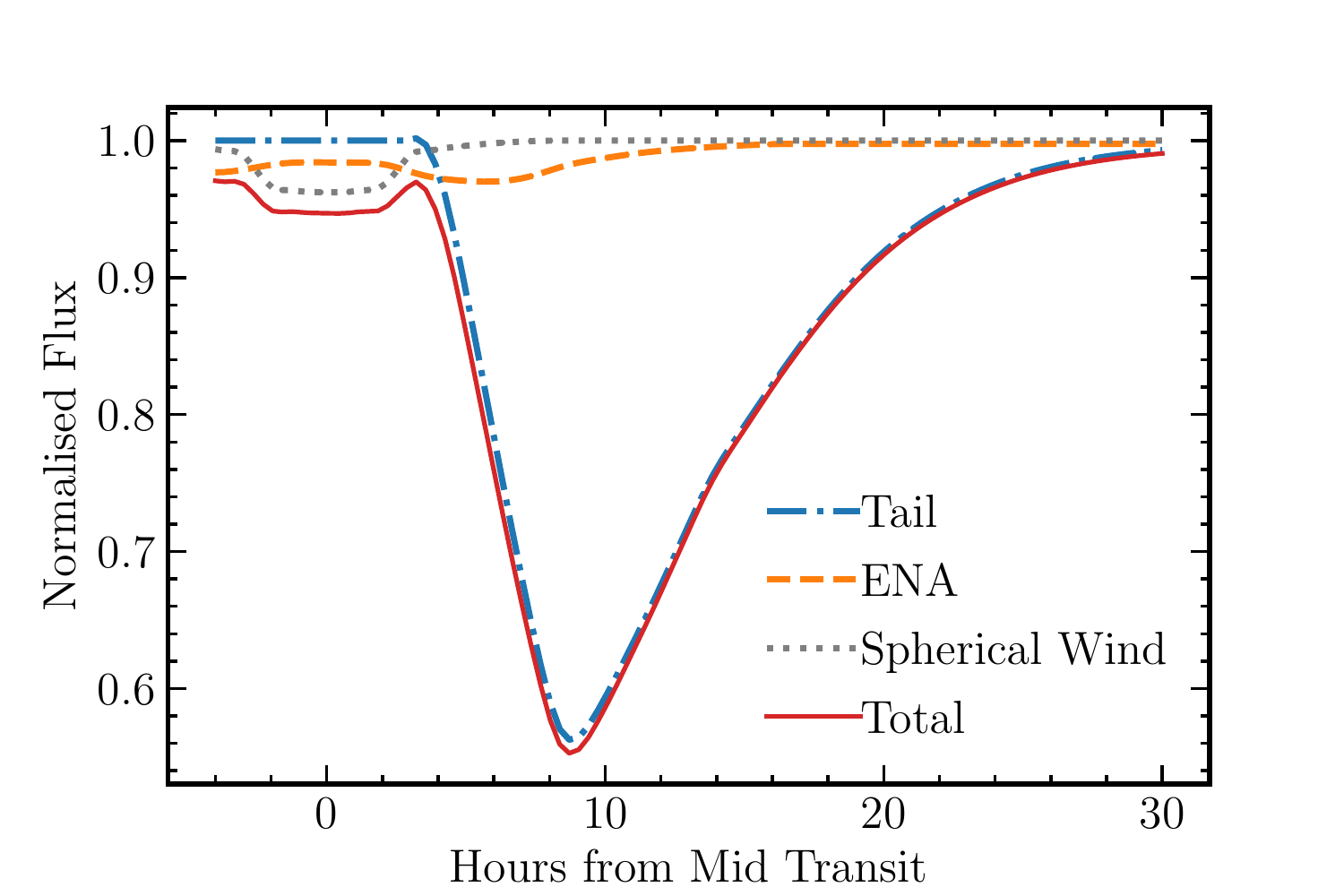}
    \caption{Light curves computed from our tail model for our sub-Neptune model (top) and hot Jupiter model (bottom) computed using ray-tracing over the stellar disc. Both planets reside at 0.15 AU from a Sun-like star. The light-curve is computed by averaging the optical depth between -50 and -150 km~s$^{-1}$. Each model show's the contribution from the spherical wind inside the planet's Hill sphere, the cometary tail and ENAs. The sub-Neptune shows a characteristic cometary tail light-curve. The hot Jupiter model shows a initial primary transit followed by a deeper transit 10 hours after the planet has transited. This second transit arises due to the time it takes to accelerate the hot Jupiter's tail into the blue-wing of the Lyman-$\alpha$ line.}
    \label{fig:light_curve}
\end{figure}

In order to determine whether our model provides a qualitative description of real Lyman-$\alpha$ lightcurves we perform ray-tracing calculation of two tail models. This ray-tracing is performed over the stellar disc for a single sub-Neptune and hot Jupiter model, both of which reside at 0.15~AU from a Sun-like star. To perform the calculation we simply assume the density is constant in each individual elliptical cross-section of the cylindrical tail, as such these calculations should be considered illustrative. The result of this exercise is shown in Figure~\ref{fig:light_curve}.
We see that the sub-Neptune produces the characteristic cometary tail seen in the observed Lyman-$\alpha$ transits of GJ 436b and GJ 3470 b. Alternatively, the hot Jupiter model shows a double transit signature discussed above. Material in the planet's Hill sphere is optically thick in the line wings giving rise to an initial transit, with a strong contribution from ENAs. This initial transit is likely to be larger in the blue-wing due to the ENAs, but have a small red-wing contribution. Material in the tail then gives rise to a delayed transit approximately 10 hours after the primary transit as material in the hot Jupiter's tail takes longer to accelerate into the blue-wing. We note this double transit, also discussed in the preceding section, was seen in the simulations of \citet{McCann2019} for hot Jupiters and it's reassuring our model qualitatively reproduces this signature seen in the full 3D radiation hydrodynamic simulations. This double transit should be targeted with future Lyman-$\alpha$ observations as both HD 209458 b and HD 189733b have not been post optical transit in Lyman-$\alpha$.

\section{Discussion}
\label{sec:discuss}

Motivated by observations and hydrodynamic simulations we have developed a physical framework to interpret the properties and observability of Lyman-$\alpha$ transits. We have shown that while a Lyman-$\alpha$ transit is a valuable method of determining if atmospheric escape is happening, if a transit is not observed it does not imply atmospheric escape is not happening, a fact we we shall demonstrate with a comparison to real systems. Additionally, the primary observable of Lyman-$\alpha$ transits is not mass-loss rates, rather it is the velocity at which the material is leaving the planet's vicinity (which can be constrained via a measurement of the transit duration). Thus, with a large enough sample of Lyman-$\alpha$ transits one could statistically test atmospheric escape models, as well as constrain the stellar wind properties in systems where a transit is not detected.  

\subsection{Comparison to real systems}
\begin{figure}
\centering
\includegraphics[width=\columnwidth]{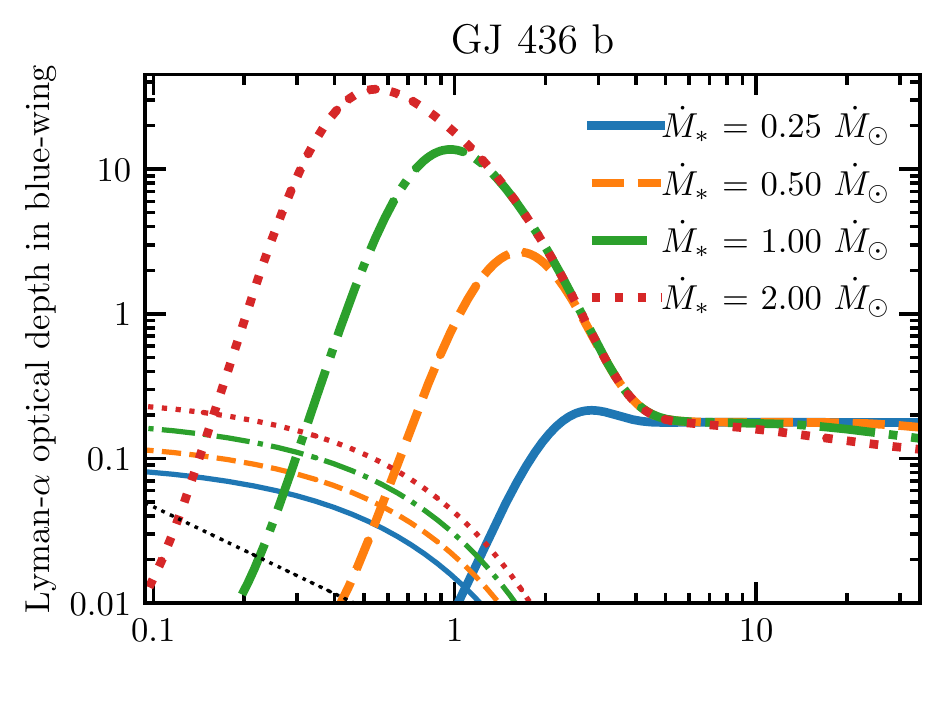}
\includegraphics[width=\columnwidth]{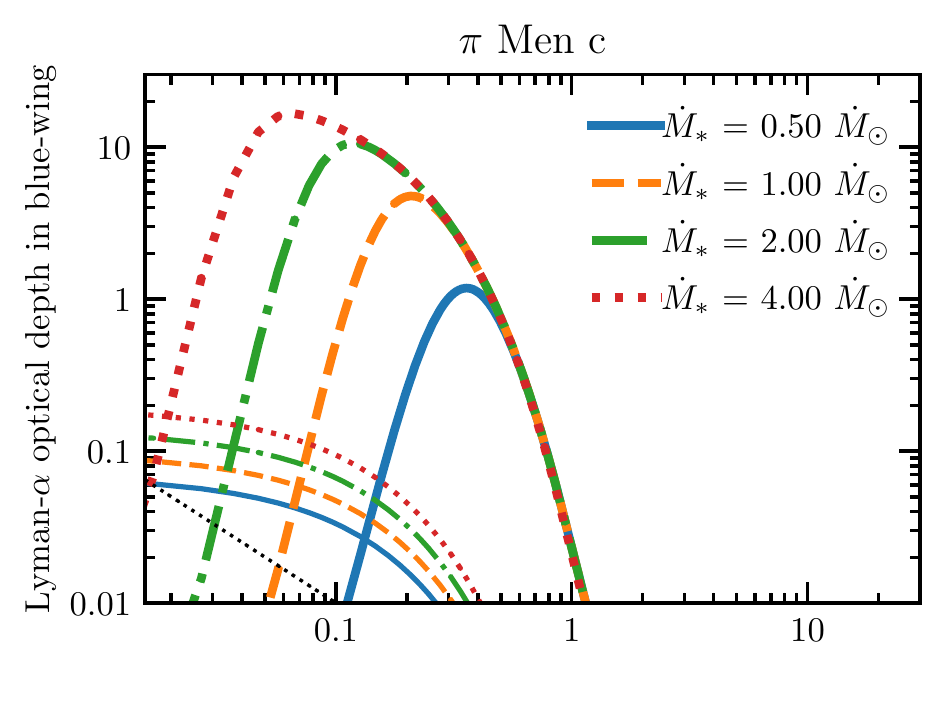}
\includegraphics[width=\columnwidth]{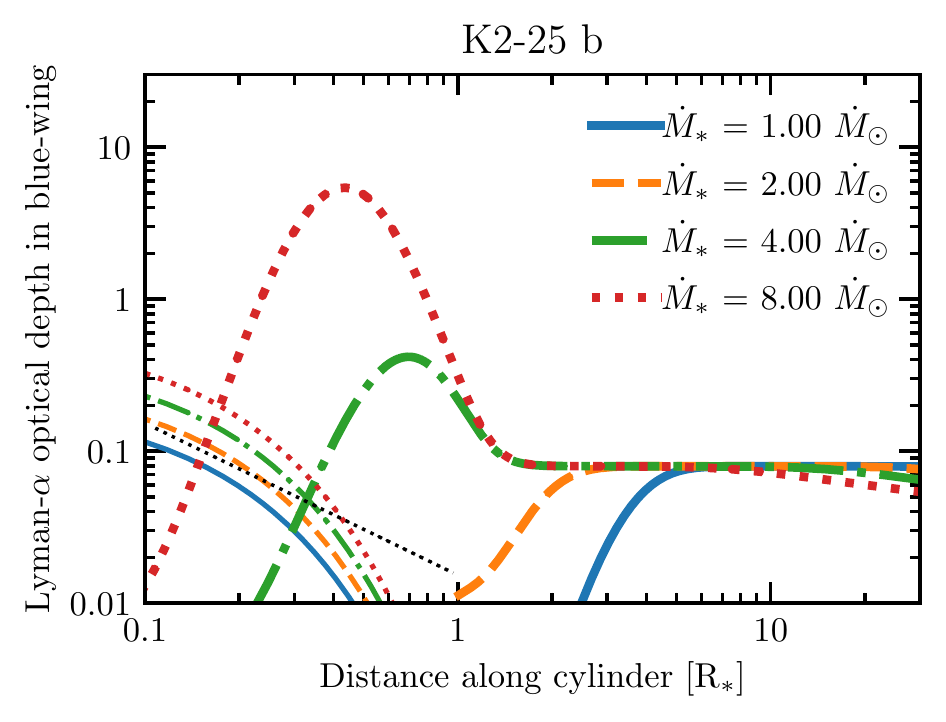}
\caption{The optical depth in the blue-wing of the Lyman-$\alpha$ line (taken to be averaged between -50 and -150~km~s$^{-1}$) as a function of distance along the cylindrical tail for GJ 436 b, $\pi$ Men c and K2-25 b. The thick lines are the contribution of planetary material in the tail, the thin lines are the contribution from ENAs. The lines indicate different stellar wind strengths compared to the solar value of $\dot{M}_\odot = 2\times10^{-14}$~erg~s$^{-2}$. The black dotted line show the optical depth arising from material in the planet's Hill sphere. These results indicate that $\pi$ Men c's and K2-25 b's non-detection can be explained by rapid photo-ionization of escaping material, while GJ 436 b can produce a long-lived neutral hydrogen tail giving rise to a large transit signature. Using the stellar-wind--X-ray flux correlation from \citet{Wood2005} we estimate wind strengths of $\sim$ 0.6, 1.9 and 1 $\dot{M}_\odot$ for GJ 436, $\pi$ Men and K2-25 respectively.}\label{fig:planet_compare}
\end{figure}
While ongoing atmospheric escape from hot Jupiters was first detected using Lyman-$\alpha$ transits \citep[e.g.][]{VidalMadjar2003,Lecavelier2010}, it is the Neptune/sub-Neptune sized planets where much of the interest lies due to escape's evolutionary role. It is these planets where the evidence is less clear cut. Knowledge of the exoplanet radius-valley allows us to identify planets that reside below it as likely terrestrial, without voluminous hydrogen dominated atmospheres. Without these hydrogen dominated atmospheres, we hypothesise hydrogen-loss from planets below the radius-valley is weak or non-existent, explaining the non-detections from Trappist-1b/c \citep{TrappistLy1}, GJ 1132b \citep{Waalkes2019}, 55 Cnc e \citep{Salz2016} and GJ 9827 b \citep{Carleo2021}. 

\begin{figure*}
\centering
\includegraphics[width=\columnwidth]{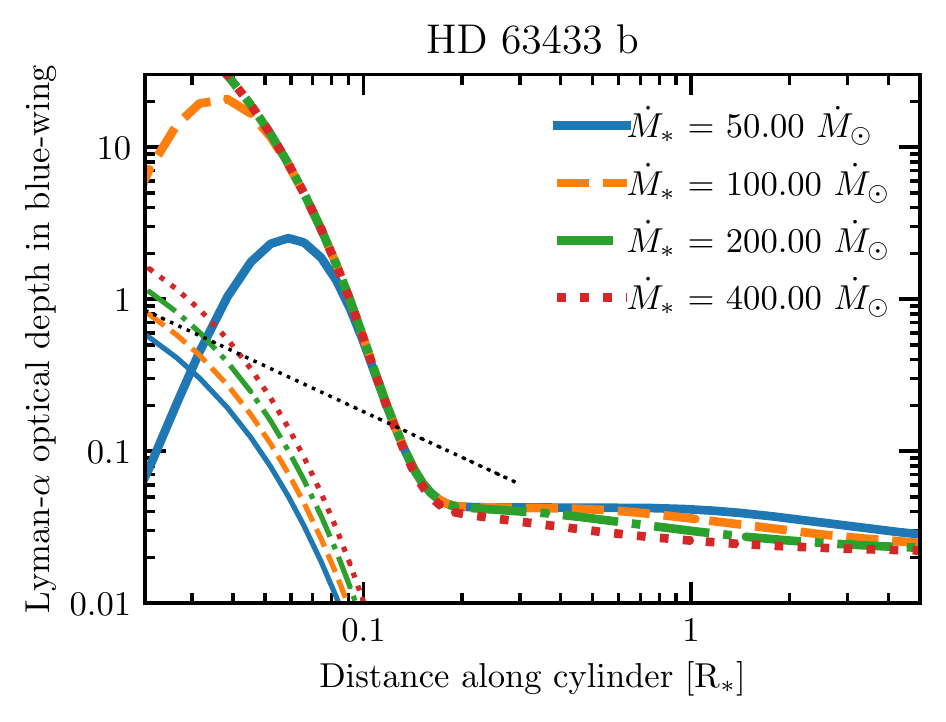}
\includegraphics[width=\columnwidth]{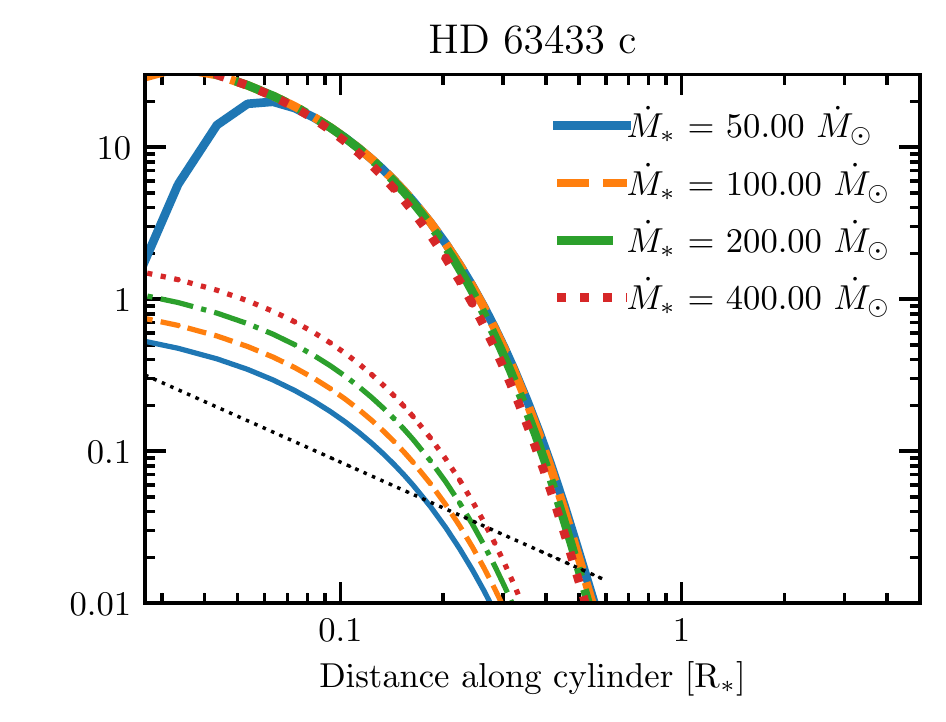}
\caption{The optical depth in the blue-wing of the Lyman-$\alpha$ line (taken to be averaged between -50 and -150~km~s$^{-1}$) as a function of distance along the cylindrical tail for the young planets HD 63433 b and c. The thick lines are the contribution of planetary material in the tail, the thin lines are the contribution from ENAs. The lines indicate different stellar wind strengths compared to the solar value of $\dot{M}_\odot = 2\times10^{-14}$~erg~s$^{-2}$. The black dotted line show the optical depth arising from material in the planet's Hill sphere. Even though these planet's are in the same system, the factor $\sim 4$ higher flux means b's tail is rapidly photo-ionized leading to no detectable transit while c's stays neutral long enough to yield a detection in the blue-wing. Using the stellar-wind--X-ray flux correlation from \citet{Wood2005} we estimate a wind strength of $\sim 200$~$\dot{M}_\odot$ for HD 63433. }\label{fig:HD63433}
\end{figure*}

What could be more puzzling are the detections around GJ 436 b \citep{Ehrenreich2015,Lavie2017}, K2-18b \citep{dosSantos2020} and HD 63433 c \citep{Zhang2021}; yet the non-detections around $\pi$ Men c \citep{GarciaMunoz2020}, K2-25 b \citep{Rockcliffe2021} and HD 63433 b \citep{Zhang2021}. These planets reside above the radius-valley and have bulk properties consistent with the presence of a hydrogen dominated atmosphere. \citet{GarciaMunoz2020} and \citet{Zhang2021} suggested that $\pi$ Men c and HD 63433 b respectively do not possess hydrogen dominated atmospheres (something that is eminently possible given its proximity to the radius-valley, \citealt{Huang2018,Gandolfi2018}). However, our model provides a more natural explanation: all planets are are undergoing atmospheric escape from hydrogen dominated atmospheres, but several are too ionized to be observable. In fact, given its youth and large size K2-25 b has the highest photoevaporation driven mass-loss rate of all, yet a {\it neutral} hydrogen tail is observable around GJ 436 but not K2-25 b. This dichotomy again highlights that Lyman-$\alpha$ transits are not primarily sensitive to mass-loss rates. To test this hypothesis explicitly we calculate the Lyman-$\alpha$ optical depths in a tail for different stellar wind parameters and nominal EUV luminosities of the three stars (GJ 436 b - $2.4\times10^{27}$~erg~s~$^{-1}$, \citealt{Youngblood2016}; $\pi$ Men c - $2.11\times10^{28}$~erg~s~$^{-1}$, \citealt{King2019}; K2-25 b - $1.32\times10^{28}$~erg~s~$^{-1}$ \citealt{Gaidos2020}). We make the choice that $\mathcal{N}_0=1$, hence our results can be considered a crude upper-limit to the expected tail lengths (but remember it only has a logarithmic dependence, Equation \ref{eqn:lya_length}). The results of these calculations are shown in Figure~\ref{fig:planet_compare}. Like our earlier models, we find ENAs dominate the initial parts of the transit for GJ 436 b and $\pi$ Men C (although the line-wings of the spherical out flow dominates for K2-25 b), yet planetary material radially accelerated by the stellar wind dominates later on. 

Our calculations indicate that provided the stellar mass-loss rate is ($\gtrsim 0.25~\dot{M}_\odot$) then we reproduce a large, long Lyman-$\alpha$ transit for GJ 436 b. While for $\pi$ Men c photo-ionization truncates the neutral hydrogen tail after a few tenths of the star's radius, consistent with the $0.24$R$_*$ upper-limit on the Lyman-$\alpha$ transit from \citet{GarciaMunoz2020}. The result for $\pi$ Men c are robust and independent of the stellar wind-loss rate. The result of K2-25 b is even more pronounced, where unless the stellar-wind mass-loss rate is $\gtrsim 8$ times than the Sun's, photo-ionization makes the tail optically thin to Lyman-$\alpha$ for all tail lengths, indicating a Lyman-$\alpha$ transit would not detect the outflow, even at arbitrarily high precision. Stellar-wind rates are higher for younger stars, and are known and theoretically expected to correlate with stellar surface X-ray flux \citep[e.g.][]{Wood2002,Johnstone2015,Blackman2016,Wood2021}, although with significant scatter \citep{Wood2021}. Comparisons with the young (300 Myr old) M-star EV Lac, that has a similar properties and a surface X-ray flux to K2-25 yields a stellar wind strength of only $\sim 1~\dot{M}_\odot$ \citep{Wood2005}. We note that \citet{Shaikhislamov2020} also explored the role of strong ionization or a weak stellar wind in making $\pi$ Men c's outflow undetectable in Lyman-$\alpha$ using simulations. {\rc They found that their simulations would provide a detectable transit  unless the ionizing flux was larger than estimated observationally (for a Solar-like stellar wind) or the that the stellar wind was weaker than Solar (for the estimated ionizing flux). Our calculations for $\pi$ Men c also indicate an undetectable transit due to ionization of the outflow for a weaker than Solar stellar wind}, but predict signatures that are very close to the constraints from current observations. Thus, both future observations to improve the detection limits and more guided simulations are warranted to explore whether $\pi$ Men c is either an evaporating hydrogen-rich planet with a highly ionized outflow, or a hydrogen poor sub-Neptune. Our model framework presented here can be used to guide more targeted simulations of $\pi$ Men c to explore this in more detail.  
We also note that we reproduce a short $\lesssim 0.5$R$_*$ cylinder length for HD 97658 b, consistent with its non-detection; however, as discussed in \citet{HD97658Ly1} this non-detection was consistent with modelling due to insufficient mass-loss (i.e $\mathcal{T}_P<1$).  Thus, even though these planets are not too dissimilar, the fact one will gives rise to a large transit and the others are non-detections highlights the sensitivity of Lyman-$\alpha$ transits to  both stellar and planetary properties. 

Finally, it is worth investigating the planets in the young 440 Myr old system, HD 63433. \citet{Zhang2021} observed an $\sim$ 11\% transit for planet c, whereas they placed a fairly stringent (2$\sigma$) upper-limit of a 3\% transit of planet b. \citet{Zhang2021} suggested that planet b may have already lost its hydrogen envelope (making it unusually large compared to its position with respect to the radius-valley)  or it has a volatile-rich but non-hydrogen dominated atmosphere (like water). Again, these suggestions are certainly viable, given the current data. However, like the case of $\pi$ Men c and K2-25 b, our model provides a simple explanation: both are hydrogen dominated atmospheres losing mass, but only the more weakly irradiated planet c has a tail that is neutral enough to give rise to a Lyman-$\alpha$ transit. Using an EUV-flux value for the star of $2.57\times10^{29}$~erg~s$^{-1}$~cm$^{-2}$ from \citet{Zhang2021} we calculate the expected Lyman-$\alpha$ tail-lengths. For planets b and c, no mass has been measured, thus we take the typical sub-Neptune mass of 5~M$_\oplus$ \citep[e.g.][]{Rogers2021} (noting that our conclusions are rather insensitive to this assumption for sub-Neptune masses in the range $\sim 2-20~$M$_\oplus$).  Our results are shown in Figure~\ref{fig:HD63433}, where they indicate we would expect no detectable transit from b (at the $<3$\% level) and a $\sim 10$\% transit from planet c.  HD 63433 is a young Sun-like (0.99 M$_\odot$) star, and scaling the X-ray luminosity from \citet{Zhang2021} to a stellar wind rate using the \citet{Wood2005} or \citet{Johnstone2015} yields mass-loss rates in the range of $\sim 75-350~\dot{M}_\odot$, and experiments indicate we only need stellar wind rates $\gtrsim 10$~$\dot{M}_\odot$ to explain the observations. Thus, in these models, the powerful stellar wind accelerated planetary material almost always dominates the Lyman-$\alpha$ optical depth over ENAs and the spherical planetary wind. While our tail models for HD 63433 c and $\pi$ Men c look similar the Lyman-$\alpha$ transit observations for HD 63433 c are about twice as sensitive. In addition, $\pi$ Men c's closer proximity to its host star means the tail height ($R_v$) is a factor of 3.3 times smaller. This again highlights that with more sensitive observations a Lyman-$\alpha$ transit could be detectable for $\pi$ Men c if it possess a hydrogen-rich atmosphere.  

While it is reassuring our model can explain the current Lyman-$\alpha$ detections and non-detections for planets above the radius-valley. We caution that these results should be considered informative, but not a quantitative comparison. For example, we have fixed the stellar wind velocity at 150~km~s$^{-1}$ and can trade stellar wind velocity for stellar mass-loss. We also did not vary the EUV luminosity within the observed uncertainties. In addition, we used a fixed mass-loss efficiency of $0.1$ and did not consider the impact of material in the Hill sphere quantitatively. A more detailed comparison will be left to future work. However, it is apparent that one does not need to invoke non-hydrogen dominated atmospheres if no Lyman-$\alpha$ transit is observed, rather a more natural explanation is perhaps that the escaping hydrogen is too ionized to be observed in Lyman-$\alpha$.

\subsection{What do Lyman-$\alpha$ transits tell us about atmospheric escape?}
In developing a simple model of Lyman-$\alpha$ transits we have gained insights into how they can probe exoplanets undergoing atmospheric escape and what physics the observations constrain. Rather confusingly at first thought, stronger EUV irradiation levels, which drive stronger mass-loss in the photoevaporation model, often lead to weaker Lyman-$\alpha$ transits. However, this is now easier to understand: while stronger mass-loss leads to more material leaving the planet, the higher photoionization rates mean the gas is more ionizied. The effect of photoionization more than counteracts the stronger mass-loss. Furthermore, we use the model to show that while the transit depth potentially encodes information about the thermodynamics/kinematics of the outflow, the star's tidal gravitational field plays a dominant role. It is, however, the transit duration that encodes important physics of atmospheric escape. Ultimately, the transit duration probes the typical distance a neutral hydrogen atom travels before it is photoionized. Thus, with an accurate measure of the EUV-flux impinging on the material lost from the planet one could determine the velocity at which the planet is ejecting material. This fact offers an exciting possibility of directly testing different atmospheric escape models, which predict quite different velocities. For example, core-powered mass-loss would predict a velocity in the range $1-2$~km~s$^{-1}$ \citep[e.g.][]{Ginzburg2018}, photoevaporation around $10$~km~s$^{-1}$ \citep[e.g.][]{Yelle2004,MurrayClay2009,Owen2012} and MHD-driven winds of order 50~km~s$^{-1}$ \citep[e.g.][]{Tanaka2014,Bourrier2016}. One of the complications is that EUV flux cannot be directly measured and must be inferred from models, which gives rise to uncertainty and an added complication. However, recent work has shown that, through detailed stellar modelling, it is possible to constrain the stellar EUV output \citep[e.g.][]{King2018,Peacock2019}. 

Our model has revealed a downside of Lyman-$\alpha$ transits is that mass-loss rates are not a primary observable; however, no other probe cleanly measures the mass-loss rate (e.g. the HeI line is degenerate; \citealt{Lampon2021}). However, here we speculate that using velocity resolved transits one may be able to extract a stronger constraint on the mass-loss rate. This possibility arises as the tail is continually being accelerated radially away from the star as it leaves the planet. Obviously the radial velocity the gas in the tail achieves after some time depends on its inertia, or more precisely the mass-per-unit-length of the tail. A direct measurement of both the bulk velocity of the gas moving along the tail, and its mass-per-unit-length would allow the mass-loss rate to be estimated. What is not clear at this stage is how accurately one needs to know the properties of the stellar wind (or Lyman-$\alpha$ luminosity if radiation pressure is the acceleration mechanism) to do this. In principle if the tail is accelerated to terminal velocity at late times in the transit then the stellar wind-speed could also be measured. Thus, it is possible velocity resolved Lyman-$\alpha$ transits allow the exciting possibility of observationally measuring both planetary and stellar mass-loss, though such an analysis must be left to future work.

\subsection{Model limitations and future work}
\label{sec:future}
The goal of the work presented here was to develop a {\rc toy} theoretical framework in which to understand Lyman-$\alpha$ observations and radiation-hydrodynamic simulations. However, there are several limitations that must be addressed in future work before we can actually use this model to quantitatively compare to observations. Firstly, we have only treated material in the planet's Hill sphere very crudely. We have shown that if the material is mainly neutral, it can present non-negligible contribution to the Lyman-$\alpha$ transit, especially for the more massive planets, like hot Jupiters, where the Hill-sphere can be comparable in size to the star. However, for the young planets considered here (K2-25 b and HD 63433 b/c) the assumption of a completely neutral Hill sphere is likely to break down. Secondly, we have assumed that the tail remains at the same orbital separation as the planet, with a constant velocity down the tail. The tail can obtain velocity of several hundred km~s$^{-1}$ while remaining neutral and optically thick over a distance of several solar radii. Over this distance the tail could move radially out several hundredths of an astronomical unit, this curvature will in turn allow the stellar wind to accelerate material down the tail. {\rc Thus, our simplified geometry of approximating the tail as trailing maximises the acceleration from the stellar wind, and will over estimate the radial velocities the material attains.} While less significant for those planets outside $\gtrsim 0.1$~AU, it will be an important correction to our model for closer in planets like GJ 436 b. This radial movement is evidenced by the particle simulations of \citet{Bourrier2016}, which shows the tail extends out to larger semi-major axis. Additionally, we made a simple estimate of the cylinder's height considering only particle dynamics, a more complete treatment would consider pressure, and its interaction with the stellar wind. All these limitations can still be addressed within the framework we have presented here without resorting, but still informed by, large scale simulations \citep[e.g.][]{Bourrier2013,CarrollNellenback2017,McCann2019,MacLeod2021,Carolan2021}. In the future, we plan to develop a fully physically consistent model with which we can quickly calculate synthetic Lyman-$\alpha$ transits, allowing us to fit the observations. {\rc Finally, it is worth discussing our approach to charge-exchange. Our approach, was to assume a collisional mixing region in a single-fluid approximation. When this approach has been adopted in simulations \citep[e.g.][]{Esquivel2019, Debrecht2022} it typically results in small ENA Lyman-$\alpha$ optical depths, as we find here. \citet{Lavie2017} adopt a particle description and \citet{Khodachenko2019} adopt a multi-fluid approach finding larger ENA Lyman-$\alpha$ optical depths. These discrepancies resulted in disagreements as to the role of ENAs in generating the observed transit signatures. While a full Monte-Carlo approach to the entire problem is not computationally feasible, more work is warranted developing an approach to ENAs that can be incorporated into both hydrodynamical simulations and simpler modelling. }

\section{Summary}
In this work we have developed the minimal framework to describe and interpret Lyman-$\alpha$ transits arising from exoplanets undergoing atmospheric escape. As material leaves the planet, it is shaped into a cometary tail {\rc (which in our simple approach, we model as a trailing tail)}, progressively photoionized by the stellar EUV radiation and accelerated radially away from the star. The initial interaction between the planetary and stellar wind can result in sufficient Lyman-$\alpha$ optical depth from {\rc planetary wind material inside the Hill sphere due to optically thick line wings and ENAs produced through charge exchange, giving rise to a transit signature}, especially for larger (hot Jupiter like) planets. Though, in general, an observable Lyman-$\alpha$ transit requires the planetary material to be radially accelerated such that its absorption can be detected in the line's blue-wing, especially for smaller (sub-Neptune like) planets. Thus, the observability of a Lyman-$\alpha$ transit is fundamentally set by the fact that the material must be sufficiently radially accelerated before it becomes too ionized. This {\rc basic} picture has allowed us to understand the detection and non-detection of Lyman-$\alpha$ transits around sub-Neptune and Neptune sized planets including GJ 436 b, K2-18 b, $\pi$ Men c, K2-25 b and HD 63433 b/c. Hence the non-detection of a Lyman-$\alpha$ transit does not necessarily mean a planet is not undergoing vigorous atmospheric escape, nor does it necessarily mean the planet does not contain a hydrogen dominated atmosphere. Our framework, will allow future targeted 3D radiation hydrodynamic simulations to systematically explore the parameter space testing the possible non-detection scenarios for sub-Neptunes: too rapidly ionized planetary material, too little planetary mass-loss (perhaps due to magnetic fields, e.g. \citealt{Owen2014}), or hydrogen poor atmospheres. 

Our framework has also allowed us to understand what properties of atmospheric escape a Lyman-$\alpha$ transit probes. The transit depth tells us about motion of the gas in the stellar tidal field and the size of the star and as such encodes weak constraints on atmospheric escape. The transit duration; however, encodes information about how the cometary tail is progressively ionized. At the most basic level it constrains the typical distance that a neutral hydrogen atom travels before it's ionized by an EUV photon. This means that with sufficiently accurate knowledge of the stellar EUV field one could extract the velocity at which the gas is ejected from the planet's Hill sphere. This velocity differs for different atmospheric escape models, indicating the exciting possibility to use Lyman-$\alpha$ transits to quantitatively test and distinguish between different atmospheric escape models with a survey. For example, the NASA MIDEX mission concept UV-SCOPE would provide such an observational capability \citep{Shkolnik2021,Shkolnik2022,UVSCOPE2022}.

We have shown that neither the transit depth or transit duration are primarily sensitive to the planetary mass-loss rate. Specifically, the transit duration only has a logarithmic dependence on planetary mass-loss rate. This means that direct mass-loss measurements are difficult to obtain from Lyman-$\alpha$ transits, and one would require an accurate model of the interaction between the planetary outflow and the circumstellar environment. We do speculate that with velocity resolved transit spectroscopy one maybe able to get a tighter constraints on the mass-loss rate by measuring the inertia of the planetary outflow as it is accelerated radially. However, what is unclear is how degenerate this is with the poorly constrained properties of stellar winds.

Thus, while Lyman-$\alpha$ transits can be used to quantitatively test atmospheric escape models. One must appeal to statistical studies on an ensemble of planets, where confounding effect of non-detections can be negated. Once a atmospheric escape model has been validated in this way, we can be confident about using it to predict mass-loss rates for evolutionary calculations. 

Extension of our minimal model to include neglected effects such as the the ionization state of hydrogen when it leaves the planet's Hill sphere and the fact the tail does not reside exactly on the planet's orbit are necessary before we can begin fitting observed Lyman-$\alpha$ transits. However, such fitting will allow us to start investigating the accuracy to which we can test atmospheric escape models using Lyman-$\alpha$ transits.

\section*{Acknowledgements}
We are grateful to the anonymous reviewers for comments which improved the manuscript. JEO is supported by a Royal Society University Research Fellowship. This project has received funding from the European Research Council (ERC) under the European Union’s Horizon 2020 research and innovation programme (Grant agreement No. 853022, PEVAP). RMC acknowledges support from NSF grant 1663706. HES gratefully acknowledges support from NASA under grant number 80NSSC21K0392 issued through the Exoplanet Research Program. AG is supported by a NASA Future Investigators in Earth and Space Science and Technology (FINESST) grant 80NSSC20K1372. We are grateful to Luca Fossati and Michael Zhang for comments on the manuscript. For the purpose of open access, the authors have applied a Creative Commons Attribution (CC-BY) licence to any Author Accepted Manuscript version arising. 

\section*{Data Availability}
The data underlying this article will be shared on reasonable request to the corresponding author.




\bibliographystyle{mnras}
\bibliography{ref} 






\bsp	
\label{lastpage}
\end{document}